\documentclass[aps,showpacs,preprintnumbers,amsmath,amssymb,nofootinbib]{revtex4}

\usepackage{graphicx}
\usepackage{color}

\def \be  {\begin{equation}}
\def \ee  {\end{equation}}
\def \ee  {\end{equation}}
\def \bea {\begin{eqnarray}}
\def \eea {\end{eqnarray}}

\begin{document}

\preprint{ECTP-2011-02}

\title{Quark-Hadron Phase Transitions in Viscous Early universe}

\author{A.~Tawfik}
\email{a.tawfik@eng.mti.edu.eg}
\email{atawfik@cern.ch}
\affiliation{Egyptian Center for Theoretical Physics (ECTP), MTI University,
 Cairo, Egypt}
\affiliation{Research Center for Einstein Physics, Freie-University Berlin, Berlin, Germany}
\author{T.~Harko}
\email{harko@hkucc.hku.hk}
\affiliation{Department of Physics and Center for Theoretical
and Computational Physics, The University of Hong Kong, Pok Fu Lam Road, Hong Kong, China}

\date{\today}

\begin{abstract}

 In the standard hot big bang theory, when the Universe was about $1-10~\mu$s old, the cosmological matter is conjectured to undergo Quantum Chromodynamics (QCD) phase transition(s) from quark matter to hadrons. In the present work, we study the cosmological quark-hadron phase transition in two different physical scenarios. First, by assuming that the phase transition would be described by an effective nucleation theory (prompt {\it first-order} phase transition), we analyze the evolution of the relevant cosmological parameters of the early Universe (energy density $\rho$, temperature $T$, Hubble parameter $H$ and the scale factor $a$) before, during and after the phase transition. To study the cosmological dynamics and the time evolution, we use both analytical and numerical methods. The case where the Universe evolved through a mixed phase with a small initial supercooling and monotonically growing hadronic bubbles is also considered in detail. The numerical estimation of the cosmological parameters, $a$ and $H$ for instance, shows that the time evolution of the Universe varies from phase to phase. As the QCD era turns to be fairly accessible in the high-energy experiments and the lattice QCD simulations, the QCD equation of state is very well defined. In light of these QCD results, we develop a systematic study of the {\it crossover } quark-hadron phase transition and  an estimation for the time evolution of the Hubble parameter during the crossover .

\end{abstract}


\maketitle


\section{Introduction}

According to the standard model of cosmology, as the Universe extremely expanded and cooled down, it is likely to expect that the cosmological background matter should undergo a series of symmetry-breaking phase transitions, at which various topological defects may have formed. The study of phase transition from quark-gluon plasma (QGP) to hadrons in the early Universe dates back to about three decades ago \cite{qgp-h1,qgp-h2,qgp-h3,qgp-h4,qgp-h5}. A first-order phase transition in various scenarios is assumed to take place \cite{ng-sze}. In one scenario, it has been suggested that QGP thermodynamically condensates into a hadron gas. In the second scenario, it is conjectured that the Universe was being supercooled and an out-of-equilibrium nucleation of hadron bubbles in the QGP surrounding should take place. In the third scenario, it has been argued that the phase transition took place in accompany with a small supercooling. Apparently, the coexistence of hadrons and QGP is accessible after the nucleation. The latter would generate fluctuations in the isothermal baryon density i.e., inhomogeneity and therefore can lead to drastic astrophysical consequences. From Yang-Mills theory, we have learned a lot about the kinetics and order of the phase transition \cite{lqcd}. The lattice QCD is a reliable method describing the strongly interacting matter for the whole temperature range starting from very low temperatures (ground state) to very high temperatures (perturbative QCD). Recently, a remarkable discovery of the QGP properties has been achieved in the heavy-ion collision program \cite{reff1,reff2,reff3,reff4}. The QGP is likely a strongly correlated phase with finite bulk and shear viscosity.

A first-order  phase transition is proceeded by bubble nucleation and rapid expansion. When at least $4-n\,$ of these bubble collide, where $n=0,1,2$, an $n$-dimensional topological defect may form in the region between them \cite{Kajantie:1986hq}. Recent lattice QCD calculations for two quark flavors suggest that QCD reliably describes a transition at $T_{c}\sim 173$ MeV \cite{TaBo07}. It is neither first- nor second-order. With increasing temperature there is a rapid change in all thermodynamic quantities. This phase transition, which could have occurred in the early universe, could lead to the formation of relic quark-gluon plasma objects, which still survive today. It will be elaborated below, that the order of the phase transition strongly depends on the mass and flavor of the quarks.

As given above, studying the first-order quark-hadron phase transition in the early Universe has a long history. It can be characterized as follows \cite{Kajantie:1986hq}. As the color deconfined QGP cools down below $T_c$, it becomes energetically favorable to form color confined hadrons (primarily the lightest Goldstone bosons; the pions and  a tiny amount of neutrons and protons, due to the conserved net baryon number). However, the new phase does not show up immediately. A characteristic feature of the first-order phase transition is that a part of the supercooling is needed to overcome the energy expense of forming the surface of the bubble and the new hadron phase. When a hadron bubble is nucleated, latent heat is released and  a spherical shock wave expands into the surrounding supercooled QGP. This reheats the plasma to the critical temperature, preventing further nucleation in a region passed by one or more shock fronts. Generally, the bubble growth is described by deflagrations with a shock front preceding the actual transition front. The nucleation stops, when the whole Universe has reheated to $T_c$. This part of the phase transition passes very fast, in about $0.05$ $\mu$sec, during which the cosmic expansion is totally negligible. After that, the hadron bubbles grow at the expense of the quark phase and eventually percolate or coalesce. When neglecting the possibility of the quark nugget production, the transition is assumed to stop, when all QGP has been converted to hadrons.

Depending on the numerical values of the parameters, both deflagrations and detonations can appear. The hadron bubbles can nucleate at very large distance scales and  the phase transition may be completed without reheating to the critical temperature. During the low temperature phase in the phase transition the bubble can grow as a supersonic deflagration consisting of a Jouguet deflagration followed by a rarefaction wave. The velocity of the supersonic deflagration varies between the sound and light velocities \cite{KuLa95}. The small-scale effects of finite wall width and surface tension have been incorporated in a numerical code, also including both the complete hydrodynamics of the problem and a phenomenological model for the microscopic entropy production mechanism at the phase transition surface \cite{KuLa96}. The decaying droplets leave behind no rarefaction wave, so that any baryon number inhomogeneity generated previously should survive the decay.

The nucleation of bubbles, the collisions of shock fronts preceding the bubble, the arrestation of the bubble growth by the reheating, the condensation of the baryon number and the resulting density perturbations after a first-order phase transition through the mixed phase have been studied in a scenario with small initial supercooling and monotonically growing hadronic bubbles \cite{Kajantie:1986hq}. The growth of bubbles after the initial nucleation event in the {\it generic} first-order cosmological phase transitions, which is characterized by the latent heat $L$, the interface tension $\sigma$ and the correlation length $\zeta $ and is driven by a scalar order parameter $\phi $ has been considered in Ref. \cite{IgKaKuLa94}. The mean distance of the nucleation $d_{\rm nuc}$ in a first-order cosmological quark-hadron phase transition has been introduced in Ref. \cite{ChMa96}. For a homogeneous nucleation $d_{\rm nuc}\leq 2 $cm. On the other hand, the impurities can lead to heterogeneous nucleation, with $d_{\rm nuc}$ of several meters. The latter value could change the outcome of the big bang nucleosynthesis. The study of the hydrodynamics of the disconnected quark regions during the final stages of the cosmological quark-hadron transition has been carried out in Ref. \cite{ReMiPa95}. It has been shown that a self-similar solution likely exists. The inclusion of the relativistic radiative transfer produces significantly different results. Furthermore, it enables the formation of high density regions at the end of the drop evaporation \cite{ReMi96}. The linear stability analysis of the relativistic detonation fronts, representing the phase interface in first-order phase transitions, has been performed in Ref. \cite{Re96}. The strong detonations are evolutionary and stable with respect to the corrugations of the front. Moreover, Chapman-Jouguet detonations appear to be unconditionally linearly stable. Taking into account the simultaneous effects of the baryon number flux suppression at the phase interface, the entropy extraction by means of the particles having long mean free paths and baryon diffusion shows that significant baryon number concentrations, up to densities above that of nuclear matter, represent an inevitable outcome within this scenario \cite{Re96}.

The abundance and size distribution of the quark nuggets formed a few microseconds after the big bang      due to a first-order QCD phase transition have been estimated in Ref. \cite{ind00}. The evolution and the collision of slow-moving true vacuum bubbles are examined in Ref. \cite{DaLi00}. The comoving  bubble walls prevent the formation of extra defects and  may lead to an increase of any primordial magnetic field. Within an effective model of QCD, the quark-hadron phase transition was studied in Ref. \cite{BoCoMa00}. In a reasonable range of the parameters of the model, bodies with a quark content between $10^{-2}$ and $10$ $M_{\odot}$ could have been formed in the early universe. A significant amount of entropy is released during the transition. The density fluctuations amplified by the vanishing sound velocity effect during the quark-hadron phase transition could lead to QGP lumps decoupled from the expansion, which rapidly transform to quark nuggets \cite{KiLee01}. The inhomogeneous nucleation, as a new mechanism for the cosmological QCD phase transition, was proposed  by Ignatius and Schwartz \cite{IgSc01}. In this model the typical distance between bubble centers is of the order of a few meters. The resulting baryon inhomogeneities may affect the primordial nucleosynthesis.

Recent lattice QCD simulations turn to be able to provide an accurate tool to study - among others - the thermodynamics of the strongly interacting matter. The critical temperature $T_c$  was a subject of different lattice QCD simulations~\cite{Fodor:2001au, deForcrand:2002ci, Allton:2002zi,
  D'Elia:2002gd, Karsch:2000kv, Karsch:2001cy, Gavai:2003mf}. We know so far that for two quark flavors ($n_f=2$) the transition is second-order or rapid crossover and $T_c\simeq 173\pm8\;$MeV. For $n_f=3$, we have a first-order  phase transition and $T_c\simeq 154\pm8\;$MeV. For $n_f=2+1$ i.e., two degenerate light quarks and one heavy strange quark, the transition is again crossover and $T_c\simeq 173\pm8\;$MeV. For the pure gauge theory, $T_c\simeq 271\pm2\;$~MeV and the deconfinement phase transition is first-order. In all these lattice QCD simulations, the quark masses are much heavier than their physical values. With recent computational facilities and modern algorithms, it is now possible to use values very close to the physical masses. This raised the critical temperature, for instance, $T_c\simeq 200~$MeV for $n_f=2+1$. From this discussion, we conclude that the order of the phase transition can be either continuous or discontinuous. It depends - among others - on the quark flavors and their masses. The extreme conditions in the early universe, like high temperatures, high densities and out-of-thermal and out-of-chemical equilibrium, likely affect the properties of the partonic matter. Yet, we have no access to study this issue. Recent lattice QCD outputs have been used in \cite{Bon} to work out the expansion law of the Universe during the cosmological quark-hadron transition. The cosmological behavior found using lattice data was compared with the one obtainable in case the transitions were first-order .  The differences between these two scenarios are too small to be tested with cosmological data, but the coming of the era of precision cosmology might open the possibility of testing the nature of the QCD transition by using cosmological data.

In the present work, we consider two cases. First, we assume that the phase transition is of first-order. The cosmological evolutions during the quark and hadron phases are investigated in detail. The main cosmological parameters are obtained for each phase. The hadron fraction $h$, whose time evolution describes the conversion process, is an important parameter to describe the phase transition and  its expression is obtained in an analytical form. $h$ seems to behave as an order parameter. The second part of this study is devoted to an extension of previous works~\cite{twfk1,twfk2,twfk3,twfk4,twfk5,twfk6,twfk7}, in which we have applied the equations of state deduced from recent lattice QCD simulations at {\it almost} physical masses and more accurate lattice configurations in order to study the cosmology of the early universe. With the use of these equations of state we can study the evolution equations of the main physical parameters of the cosmological models. In light of these QCD results, we develop a systematic study of the {\it crossover } quark-hadron phase transition and  an estimation for the time evolution of the Hubble parameter during the crossover in the presence of bulk viscous effects.

This paper is organized in the following manner. In Section~\ref{field}, the background geometry and the gravitational field equations are written down and  the description of the viscous effects in different theoretical models is presented. In Section~\ref{SecII}, we lay down the equations of state and the relevant physical quantities, necessary for the discussion of the first-order quark-hadron phase transition. In Section~\ref{SecIII} we analyze in detail the dynamics of the Universe during first-order  quark-hadron phase transition. The phase transition in the lattice QCD simulations and the heavy-ion collisions and the QCD equation of state (EoS) are discussed  in Section~\ref{qgpEoS}. The cosmological evolution of the Universe during the crossover in the presence of bulk viscous effects is analyzed in Section~\ref{cosm}. The cosmological implications of our results are discussed in Section~\ref{cosm1}.  We discuss and summarize our results in Section \ref{SecV}.

In the present paper we use natural units with $c=\hbar =k_B=1$, in which $8\pi G=1/m_{Pl}^2=1.687\times 10^{-43}\;{\rm MeV^{-2}}$, where $m_{Pl}$ is the ''reduced'' Planck mass. The "reduced" Planck time is given by $t_{Pl}=1/m_{Pl}= 4.0\times 1152\times10^{-22}$ MeV$^{-1}$.

\section{Geometry and field equations}\label{field}

We assume that the  early Universe is filled with a bulk viscous
cosmological fluid and  its geometry is given by a spatially flat Friedmann-Lemaitre-Robertson-Walker (FLRW) metric
\begin{equation}  \label{1}
ds^{2}=dt^{2}-a^{2}(t) \left[ dr^{2}+r^{2}\left(d\theta
^{2}+\sin ^{2}\theta d\phi ^{2}\right) \right] ,
\end{equation}
where $a(t)$ is the dimensionless scale factor, which describes the expansion of the universe.
At a vanishing cosmological constant, the Einstein gravitational field equations in the flat Universe read
\begin{equation}
R_{ik}-\frac{1}{2}g_{ik}\, R=\frac{1}{m_{Pl}^2}\, T_{ik}.  \label{ein}
\end{equation}
 The energy-momentum tensor of the bulk viscous cosmological fluid filling the very early Universe is given by \cite{Ma95}
\begin{equation}
T_{i}^{k}=\left(  \rho +p+\Pi\right)  u_{i}u^{k}-\left(  p+\Pi\right)
\delta_{i}^{k},\label{1_a}%
\end{equation}
where indices $i, k$ take discrete values $0,1,2,3$, $\rho$ is the energy density, $p$ is the thermodynamic pressure, $\Pi $ is the bulk viscous pressure and  $u_{i}$ is the four velocity, satisfying the normalization condition $u_{i}u^{i}=1$. The particle and entropy fluxes are defined according to $N^{i}=nu^{i}$ and $S^{i}=sN^{i}-\left(  \tau\Pi^{2}/2\xi T\right) u^{i}$, where $n$ is the number density, $s$ is the specific entropy, $T$ is the finite temperature, $\xi$ is the bulk viscosity coefficient and  $\tau$ gives the relaxation coefficient for the transient bulk viscous effect (i.e. the relaxation time), respectively. The evolution of the cosmological fluid is subject to obeying the dynamical laws of the particle number conservation $N_{\; ;i}^{i}=0$ and Gibbs' equation $T d\rho=d\left(\rho/n\right)+p d\left(1/n\right)$ \cite{Ma95}. In the following, we shall also suppose that the energy-momentum tensor of the cosmological fluid is conserved, i.e., $T_{i;k}^{k}=0$, where $;$ denotes the covariant derivative with respect to the metric.

The bulk viscous effects can generally be described by means of an effective
pressure $\Pi $, formally included in the effective thermodynamic pressure $%
p_{eff}=p+\Pi $ \cite{Ma95}. Then in the comoving  frame the energy-momentum tensor
has the components $T_{0}^{0}=\rho ,T_{1}^{1}=T_{2}^{2}=T_{3}^{3}=-p_{eff}$.
For the line element given by Eq.~(\ref{1}), the Einstein field equations read
\begin{eqnarray}  \label{dH}
H^{2} &=& \frac{1}{3m_{Pl}^2} \;\rho, \\
\dot H + H^2 &=& -\frac{1}{6m_{Pl}^2} \; \left( 3p_{eff}+\rho \right),
\label{drho}
\end{eqnarray}
where one dot denotes the derivative with respect to the time $t$, $G$ is the gravitational constant and $H(t)=\dot a(t)/a(t)$ is the Hubble parameter. Expressions (\ref{dH}) and (\ref{drho}) lead to a {\it generic} expression for the time evolution of $H$:
\bea
\dot H &=& -\frac{1}{2m_{Pl}^2} (\rho + p + \Pi). \label{Eck-Hh0}
\eea
From the field equations or with the use of the conservation of the energy-momentum tensor we obtain the following equation (Bianchi identity), relating the time variation of the energy density to the Hubble parameter:
\be\label{drho2}
\dot{\rho }+3\left(\rho +p_{eff}\right)H=0.
\ee
In order to solve the field equations, we necessarily need an equation of state and an estimation for the bulk viscous $\Pi$, characterizing the viscous properties of the matter in the expanding universe.

\subsection{Eckart relativistic viscous fluid}

The first attempts at creating a theory of relativistic fluids were those of Eckart \cite{Ec40} and Landau and Lifshitz \cite{LaLi87}. These theories are now known to be pathological in several respects. Regardless of the choice of the equation of state, all equilibrium states in these theories are unstable and in addition signals may be propagated through the fluid at velocities exceeding the speed of light $c$ violating the causality principle. These problems arise due to the nature of the first-order  of this theory [review Eq. (\ref{eckart_s})], that it considers only the first-order  deviations from the equilibrium leading to parabolic differential equations, because of the infinite speeds of propagation for the heat flow and viscosity, which contradicts the principle of causality. Conventional theory is thus applicable only to phenomena which are quasistationary, i.e., slowly varying on space- and time-scales characterized by mean free path and mean collision time.

The Eckart theory can be applied on modelling the cosmic background fluid as a continuum with a well-defined average 4-velocity field $u^{\alpha}$ where $u^{\alpha}u_{\alpha}=-1$. The vector number density $n^{\alpha}=n\, u^{\alpha}$ can be estimated, when unbalanced creation/annihilation processes take place; $n^{\alpha}_{;\alpha}=0$. This apparently means that
\begin{equation}
 \dot n + 3\, H\, n = 0,
\end{equation}
where the Hubble parameter $H= u^{\alpha }_{;\alpha }$. In the case of a viscous fluid, the entropy current,
\begin{eqnarray} \label{eckart_s}
S^{\alpha}&=&s\, n\, u^{\alpha},
\end{eqnarray}
is no longer conserved. The covariant form of second law of thermodynamics is $S^{\alpha}_{;\alpha}\ge 0$ and the divergence of entropy current is given by $T S^{\alpha}_{;\alpha}=-3H\Pi$. This is another feature of Eckart's theory. It violates the second law of thermodynamics.

The evolution of the cosmological fluid is subject to the dynamical laws of
particle number conservation $N_{;i}^{i}=0$ and Gibbs' equation
$Td\rho=d\left(\rho/n\right)+pd\left(1/n\right)$.
 Then, from the Gibbs equation, the covariant entropy current can be obtained as
\begin{equation} \label{entr}
\Pi=-3\, \xi\, H.
\end{equation}
This is a linear {\it first-order} relationship between the thermodynamical flux $\Pi$ and the corresponding force $H$. Substituting in Eq. (\ref{Eck-Hh0})  results in
\begin{eqnarray}
\dot H &=& - \frac{1}{2m_{Pl}^2} (\rho + p - 3\, \xi\, H). \label{eq:evlu}
\label{Eck-Hh}
\end{eqnarray}

\subsection{Israel-Stewart relativistic viscous fluid}

A relativistic second-order theory was introduced by Israel and Stewart \cite{Is76,IsSt76} and  further developed by  Hiscock and Lindblom \cite{HiLi89} through the {\it extended} irreversible thermodynamics. In this model, the deviations from equilibrium (bulk stress, heat flow and shear stress) are treated as independent dynamical variables, resulting in 14 dynamical fluid variables to be determined. The causal thermodynamics and its role in general relativity are reviewed in Ref. \cite{Ma95}. A general {\it algebraic} form for $S^{\alpha}$ including a {\it second-order} term in the dissipative thermodynamical flux $\Pi$ \cite{Is76,IsSt76} reads
\begin{equation}
S^{\alpha}=s\, n\, u^{\alpha}+\beta\, \Pi^2\, \frac{u^{\alpha}}{2T},
\end{equation}
where $\beta$ is a proportionality constant.

For the evolution of the bulk viscous pressure, we adopt the causal evolution
equation \cite{Ma95} obtained in the simplest way (linear in $\Pi)$ to
satisfy the $H$-theorem (i.e., for the entropy production to be nonnegative,
$S_{;i}^{i}=\Pi^{2}/\xi T\geq0$ \cite{Is76,IsSt76}). According to the causal relativistic IS theory, the evolution equation of the bulk viscous pressure reads~\cite{Ma95}
\begin{equation}  \label{8}
\tau \dot{\Pi}+\Pi =-3\, \xi\, H-\frac{1}{2}\ \tau\,  \Pi\, \left(3\, H+\frac{\dot{\tau}}{%
\tau }-\frac{\dot{\xi}}{\xi }-\frac{\dot{T}}{T}\right),
\end{equation}
where $\tau $ is the relaxation time.
In order to have a closed system from Eqs. (\ref{dH}), (\ref{drho2}) and (\ref{8}), we have to take into consideration equations of state for the pressure $p$, the temperature $T$ and  the relaxation time $\tau $, respectively.

\section{first-order quark-hadron phase transition}\label{SecII}

In this Section, we outline the relevant thermodynamic quantities of
the quark-hadron phase transition, which will be used in the
following sections. Note that the scale of the cosmological QCD transition is given by
the Hubble radius $R_{H}$ at the transition: $R_{H}\sim
m_{Pl}/T_{c}^{2}\sim 10\,$km, where $T_c$ is the critical
temperature. The mass inside the Hubble volume is $\sim
1\,M_{\odot}$. The expansion time scale is $10^{-5}$ s, which
should be compared with the time-scale of QCD, $1$ fm/c$\simeq
10^{-23}$ s. Even the rate of the weak interactions exceeds the
Hubble rate by a factor of $10^{7}$. Therefore, in this phase
the photons, the leptons, the quarks and the gluons (or pions) are lightly coupled
and may be described as a single, adiabatically expanding fluid
\cite{ChMa96}.

At high temperatures $T>T_{c}$, the baryon number density $n_{B}$ may be defined as $n_{B}=\left( 1/3\right) \sum \left(n_{q}-n_{\bar{q}}\right) $, where $n_{q}\left( n_{\bar{q}}\right)$ is the number density of a specific quark (anti-quark) flavor and  the sum is taken over all quark flavors. In utilizing these relations, it is apparent that QGP matter is assumed to be characterized as an ideal gas. At $T<1$ GeV only the $u$, $d$ and $s$ quarks contribute significantly. At low temperatures $T<T_{c}$ the baryon number density is defined as $n_{B}=\sum \left( n_{b}-n_{\bar{b}}\right) $, with the summation extended over all baryon species $b$.
In order to study the quark-hadron phase transition it is
necessary to specify EoS of the matter, in both
quark and hadron state. Giving an equation of state is equivalent
to give the pressure as a function of the temperature $T$ and
chemical potential $\mu$.

At high temperatures the quark chemical potentials are equal,
because of the weak interactions which apparently keep them in chemical equilibrium and 
the chemical potentials for leptons are assumed to vanish. Thus
the chemical potential for a baryon is defined by $\mu _{B}=3\mu
_{q}$. The baryon number density of an ideal Fermi gas of three
quark flavors is given by $n_{B}\simeq T^{2}\mu _{B}/3$, leading
to $\mu _{B}/T\sim 10^{-9}$ at $T>T_{c}$. At low temperatures $\mu
_{B}/T\sim 10^{-2}$. Therefore the assumption of a vanishing
chemical potential at the phase transition temperature in both
quark and hadron phase represents an excellent approximation for
the study of EoS of the cosmological matter in
the early universe. In addition to the strongly interacting matter
we assume that in each phase there are present leptons and
relativistic photons, satisfying equations of state similar to
that of hadronic matter \cite{Kajantie:1986hq}.

\subsection{Thermodynamic parameters of the quark and hadronic matter}

The equation of state of the ideal gas in QGP phase can generally be given in the form
\begin{eqnarray}
\rho_{q} &=& 3\, a_{q}T^{4} + V(T), \\
p_{q} &=& a_{q}\, T^{4} - V(T),
\label{eqq}
\end{eqnarray}
where $V\left(T\right)$ is the self-interaction potential. $a_{q}=\left(\pi^{2}/90\right) g_{q}$, with $g_{q}=16+(21/2)N_{F}+14.25=51.25$ and $N_{F}=2$. As given in Ref. \cite {BoCoMa00}, the self-interaction potential reads
\begin{equation}
V\left( T\right) =B+\gamma _{T}T^{2}-\alpha _{T}\, T^{4},  \label{V}
\end{equation}
where $B$ is the bag constant, $\alpha _{T}=7\pi ^{2}/20$ and 
$\gamma_{T}=m_{s}^{2}/4$, with $m_{s}$ is the mass of the strange
quark  $\in \left( 60-200\right) $ MeV. The form
of the potential $V$ corresponds to a physical model in which the
quark fields are interacting with a chiral field formed with the
$\pi $ meson field and a scalar field. If the temperature effects
can be ignored, EoS in the quark phase takes the
form of the MIT bag model equation of state, $p_{q}=(\rho
_{q}-4B)/3$, MIT stands for Massachusetts Institute for Technology. The results obtained in the low energy hadron spectroscopy, the
heavy-ion collisions and the phenomenological fits of the light hadron
properties give an estimation for $B^{1/4}$. It ranges between $100$ and $200$ MeV
\cite{LePa92}.

In the hadron phase, we assume that the cosmological fluid is consisting of
an ideal gas of massless pions and nucleons described by the
Maxwell-Boltzmann statistics. The energy density $\rho_{h}$ and
pressure $p_{h}$ can be respectively approximated by
\begin{eqnarray}
p_{h}\left( T\right) &=&\frac{1}{3}\rho _{h}\left( T\right) =a_{\pi }T^{4},
\label{eqh}
\end{eqnarray}
where $a_{\pi }=\left(\pi^{2}/90\right) g_{h}$ and
$g_{h}=17.25$. For the entropy densities $s(T)=dp/dT$ in the two phases we obtain
\begin{eqnarray}
s_{q}(T)&=&-2\gamma _{T}T+4\left( a_{q}+\alpha _{T}\right) T^{3}, \\
s_{h}\left(T\right) &=& 4a_{\pi }T^{3}.
\end{eqnarray}
The critical temperature $T_{c}$ is defined by the condition $p_{q}\left(T_{c}\right) =p_{h}\left(T_{c}\right)$ \cite{Kajantie:1986hq} and  is given, in the present model, by
\begin{equation}
T_{c}^2 = \frac{\gamma _{T}+\sqrt{\gamma _{T}^{2}+4\left(
a_{q}+\alpha _{T}-a_{\pi }\right) B}}{2\left( a_{q}+\alpha
_{T}-a_{\pi }\right) }. \label{Tc}
\end{equation}

For $m_{s}=200$ MeV and $B^{1/4}=200$ MeV, the transition temperature is of the order of $T_{c}\simeq 125$ MeV. According to the first-order of the phase transition, all the physical quantities, like the energy density, pressure and entropy, exhibit discontinuities across the critical curve. At the critical temperature, the ratios of the relevant physical quantities, the energy and the entropy density, respectively, are given by
\begin{equation}
\frac{\rho _{q}\left( T_{c}\right) }{\rho _{h}\left( T_{c}\right) }=\frac{%
4a_{q}T_{c}^{4}-p_{q}\left( T_{c}\right) }{3a_{\pi }T_{c}^{4}}=\frac{%
4a_{q}T_{c}^{4}-p_{h}\left( T_{c}\right) }{3a_{\pi }T_{c}^{4}}=\frac{%
4a_{q}-a_{\pi }}{3a_{\pi }},
\label{eq_ratiorho} 
\end{equation}
and
\begin{eqnarray}
\frac{s_{q}\left( T_{c}\right) }{s_{h}\left( T_{c}\right)} 
&=& \frac{\gamma _{T}a_{\pi }+\left( a_{q}+\alpha _{T}\right)
\sqrt{\gamma _{T}^{2}+4\left( a_{q}+\alpha _{T}-a_{\pi }\right)
B}}{a_{\pi }\left( \gamma _{T}+\sqrt{\gamma _{T}^{2}+4\left(
a_{q}+\alpha _{T}-a_{\pi }\right) B}\right) },
\label{eq_ratios}
\end{eqnarray}
respectively. For $m_{s}=200$ MeV and $B^{1/4}=200$ MeV, the ratios $\rho_{q}\left(T_{c}\right)/\rho_{h}\left(T_{c}\right)$, given by Eq.~(\ref{eq_ratiorho}) and  $s_{q}\left(T_{c}\right)/s_{h}\left(T_{c}\right)$, given by Eq.~ (\ref{eq_ratios}), equal $3.62$ and $4.628$, respectively. So far, we conclude that the energy density and entropy suddenly decrease to nearly one-fifth of its value, when the system undergoes a first-order phase transition at $T_c$. According to the first-law of thermodynamics, the entropy $s$ can be expressed in terms of the pressure $p$ and the energy density $\rho$, so that at vanishing chemical potential $\mu$, $s\,T=p+\rho$. It is apparent that the sudden decrease in $\rho $ nearly equals the decrease in $s$, at fixed $T$ and slightly changing constant $p$. If the temperature effects in the self-interaction potential $V$ are neglected, $\alpha_{T}=\gamma_{T}\simeq 0$, then from Eq.~(\ref{Tc}), we obtain the well-known relation between the critical temperature and the bag constant, $B=\left(g_{q}-g_{h}\right) \pi^{2}T_{c}^{4}/90$ \cite{Kajantie:1986hq}.

\section{Dynamics of the Universe during the quark-hadron phase transition}\label{SecIII}

The quantities to be traced through the quark-hadron phase transition  are the energy density $\rho $, the temperature $T$ and the scale factor $a$. These quantities are determined by the gravitational field Eqs. (\ref{dH}) and (\ref{drho2}) and by the equations of state (\ref{eqq}), (\ref{V}) and (\ref{eqh}). We shall consider now the evolution of the Universe before, during and after the phase transition.

\subsection{Cosmological evolution in the quark phase (prior to the quark-hadron phase transition)}

\begin{figure}[htb!]
\includegraphics[width=12.cm]{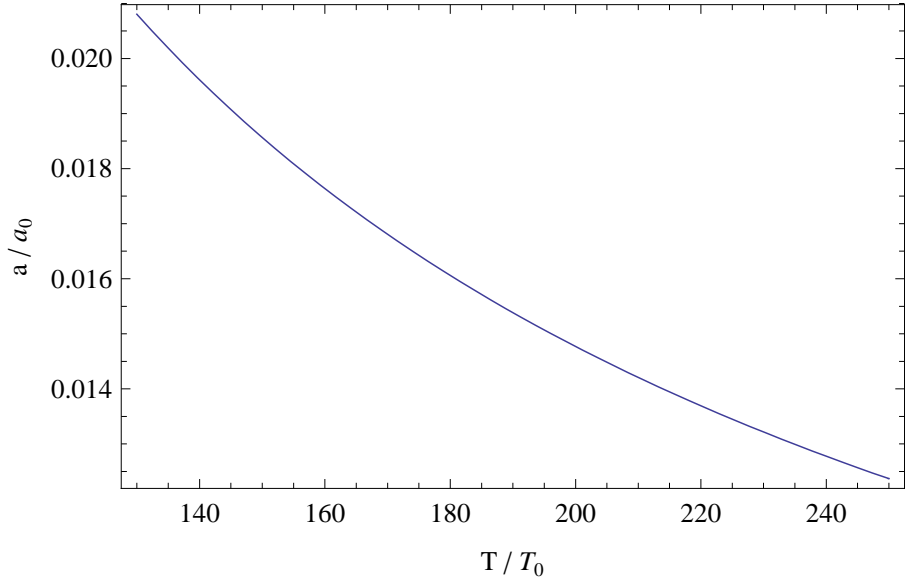}
\caption{The dependence of the scale factor $a$ of the temperature $T$ during the quark phase for $T_0=250$ MeV. The $a$-$T$ relation is almost independent on the mass $m_s$ of the strange quark. }
\label{fig1}
\end{figure}

Before the phase transition, at $T>T_{c}$, the Universe is likely in the partonic phase. With the use of the equations of state of the quark matter and  of the Bianchi identity, Eq.~(\ref{drho2}), the time evolution of the scale factor can be written in the form
\begin{equation}
H(T)=\frac{\dot{a}}{a}=-\frac{3a_{q}-\alpha_{T}}{3a_{q}}\frac{\dot{T}}{T}-\frac{1}{6}\frac{\gamma_{T}}{a_{q}}\frac{\dot{T}}{T^{3}},  \label{rdot}
\end{equation}
and can be integrated to give the following scale factor-temperature relation:
\begin{eqnarray}\label{a}
a(T) &=& a_0\left(T/T_0\right)^{\left(\alpha_{T}/3a_q-1\right)} 
\exp\left\{\frac{1}{12}\frac{\gamma_{T}}{a_qT_0^2}\left[ \left(\frac{T_0}{T}\right)^2-1\right]\right\},
\end{eqnarray}
where $a_{0}$ is the initial value of the scale factor corresponding to the temperature $T=T_0$ of the universe, $a\left(T_0\right)=a_0$. In Fig.~\ref{fig1}, the variation of the scale factor of the Universe during the quark phase is presented as a function of the temperature $T$. Because of the expansion of the Universe the temperature is decreasing with the increase in the comoving  time $t$. Therefore the scale factor $a$ increases with the decreasing $T$. The exact numerical values of $a(t)$ strongly depend on the initial $T_0$ value.

In order to have an analytical insight into the evolution of the cosmological quark matter, we consider the simple case in which the temperature corrections can be neglected in the self-interaction potential $V$. In this case $V=B={\rm constant}$ and  EoS of the quark matter is given by the bag model equation of state, $p_{q}=\left(\rho_{q}-4B\right)/3$. Thus, Eq.~(\ref{drho2}) can immediately be integrated to give the following simple scale factor-temperature relation:
\begin{equation}
\frac{a(T)}{a_0}\simeq \frac{T_{0}}{T}.
\end{equation}
Hence the presence of a temperature-dependent potential term $V(T)$ in the quark matter EoS drastically modifies the scale factor-temperature relationship. The same result can be obtained by taking $\alpha _T=\gamma _T=0$ in Eq.~(\ref{a}).

\begin{figure}[htb!]
\includegraphics[width=12.cm]{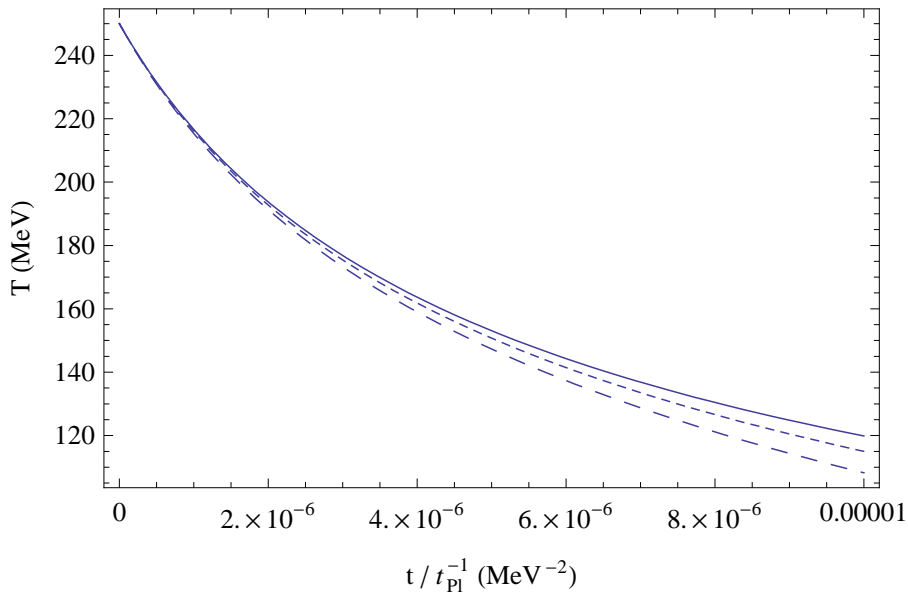}
\caption{Time dependence of the temperature of the Universe $T$ during the quark phase for a strange quark mass  $m_s=200$ MeV and different values of the bag constant: $B^{1/4}=100$ MeV (solid curve), $B^{1/4}=200$ MeV (dashed curve) and  $B^{1/4}=250$ MeV (long-dashed curve), respectively.}
\label{fig2}
\end{figure}

\begin{figure}[htb!]
\includegraphics[width=12.cm]{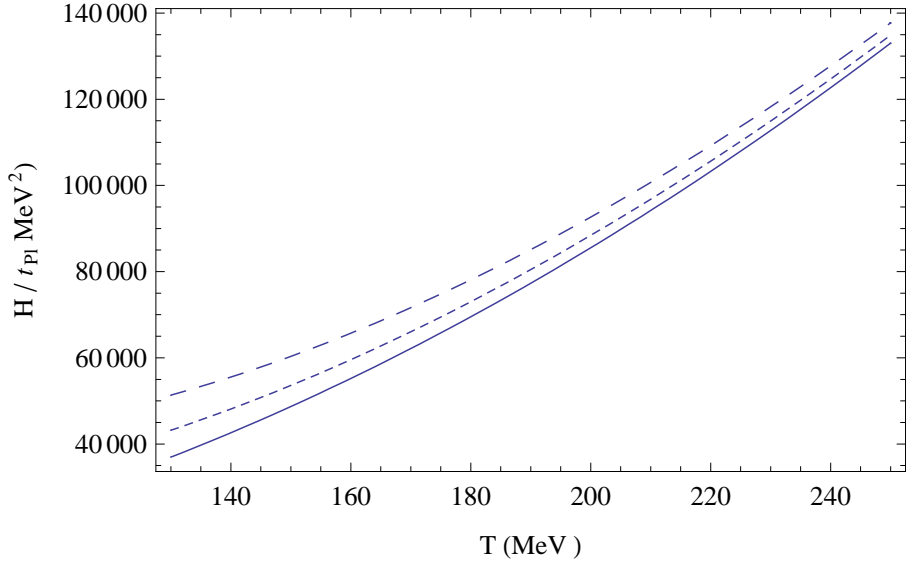}
\caption{Temperature dependence of the Hubble parameter $H$ during the quark phase for a strange quark mass of  $m_s=200$ MeV and different values of the bag constant: $B^{1/4}=100$ MeV (solid curve), $B^{1/4}=200$ MeV (dashed curve) and  $B^{1/4}=250$ MeV (long-dashed curve), respectively.}
\label{fig2b}
\end{figure}

With the use of Eq.~(\ref{rdot}) and from the gravitational field
equations, we obtain an expression describing the evolution of the
temperature of the Universe in the quark phase, given by
\begin{equation}
\frac{dT}{dt}=-\frac{T^{3}}{\sqrt{3}m_{Pl}}\frac{\sqrt{\left(3a_{q}-\alpha _T\right)T^{4}+\gamma
_{T}T^{2}+B}}{\left[ \left( 3a_{q}-\alpha_{T}\right) /3a_{q}%
\right] T^{2}+\left(\gamma_{T}/6a_{q}\right)}.
\end{equation}
The variation of the temperature in the quark phase is presented, for different values of the bag constant $B$, in Fig.~\ref{fig2}. The temperature dependence of the Hubble parameter during the quark-gluon phase is represented, for different values of the bag constant $B$, in Fig.~\ref{fig2b}.

\subsection{Cosmological dynamics during the first-order  quark-hadron phase transition}

During the quark-hadron phase transition, the temperature and the pressure are constants, $T=T_{c}$ and $p=p_{c}$, respectively. The entropy $S=s\, a^{3}$ and the enthalpy $W=\left(\rho+p\right) a^{3}$ are conserved quantities. The energy density $\rho \left(t\right)$ decreases from $\rho_{q}\left(T_{c}\right) \equiv \rho_{Q}$ to $\rho_{h}\left(T_{c}\right) \equiv \rho_{H}$. At the critical temperature $T_{c}=125$ MeV, we have $\rho_{Q}\simeq 5\times 10^{9}$ MeV$^{4}$ and $\rho_{H}\simeq 1.38\times 10^{9}$ MeV$^{4}$, respectively. The value of the pressure of the cosmological fluid during the phase transition is $p_{c}\simeq 4.6\times 10^{8}$ MeV$^{4}$. Following \cite{Kajantie:1986hq}, it is convenient to replace $\rho \left(t\right)$ by the volume fraction of matter in the hadron phase
\bea
\rho (t) &=& \rho_{H} h(t)+\rho_{Q}\left[1 - h(t)\right] = \rho_{Q} \left[1 + n h(t)\right],
\eea
where $n=\left(\rho_{H}-\rho_{Q}\right)/\rho_{Q}$ is the relative density and $t$ is the comoving  cosmological time. It is obvious that at the beginning  of the quark-hadron phase transition, the quantity $h(t_{c})$ vanishes, where $t_{c}$ is the time corresponding to the beginning of the phase transition and $\rho \left(t_{c}\right) \equiv \rho_{Q}$. At the end of the quark-hadron transition, $h\left(t_{h}\right)=1$, where $t_{h}$ is the time at which the phase transition ends  corresponding to $\rho \left(t_{h}\right) \equiv \rho_{H}$. At $t>t_{h}$, the Universe enters in the hadronic phase.

From Eq.~(\ref{drho2}), we obtain an expression for the Hubble parameter
\begin{equation}\label{vareps}
H = -\frac{1}{3}\frac{\left(\rho_{H}-\rho
_{Q}\right) \dot{h}}{\rho_{Q}+p_{c}+\left(\rho_{H}-\rho_{Q}\right)h}=-\frac{1}{3}\frac{\varepsilon \dot{h}}{1+\varepsilon h},
\end{equation}
where $\dot{h}$ denotes the time derivative of the hadron fraction parameter $h$, which can be utilized as an order parameter.
\begin{equation}
\varepsilon =\frac{\rho_{H}-\rho_{Q}}{\rho_{Q}+p_{c}}.
\end{equation}
Then, Eq.~(\ref{vareps}) immediately leads to the scale factor,
\begin{equation}\label{eq:atc1}
a(t)=a\left( t_{c}\right) \left(1+\varepsilon  h(t)\right)^{-1/3},
\end{equation}
where we have used the initial condition $h\left(t_{c}\right)=0$. The evolution of the fraction of the matter in the hadronic phase is described as
\begin{equation}
\dot h(t)=-\frac{1}{m_{Pl}}\sqrt{3 [1 + n\, h(t)]\; \rho _Q}\; \left[ h(t)+\frac{1}{\varepsilon}\right],
\end{equation}
with the general solution given by
\begin{equation} \label{eq:ht}
h(t)= \frac{n-\varepsilon }{n\varepsilon }\;
      {\rm sech^2}
      \left[\sqrt{\frac{3}{4}\,\left(1-\frac{n}{\varepsilon }\right)\rho_Q}\; \frac{t-t_c}{m_{Pl}}\right]
      - \frac{1}{\varepsilon }.
\end{equation}

As given above, the quark-hadron phase transition ends up, when the value of $h(t)$  reaches $1$. Then, the time $t_{h}$ at which the phase transition ends reads
\begin{equation}
t_h = t_c + 2 \sqrt{\frac{\varepsilon }{3 (\varepsilon -n) \rho_Q}}\;
   {\rm sech^{-1}} \left(\sqrt{\frac{n (\varepsilon +1)}{n-\varepsilon }}\right).
\end{equation}
At the end of the phase transition the scale factor of the
Universe has the value, Eq.~(\ref{eq:atc1})
\begin{equation} \label{eq:atc2}
a\left(t_{h}\right) = a\left( t_{c}\right)\left( \varepsilon +1\right) ^{-1/3}.
\end{equation}

\begin{figure}[htb!]
\includegraphics[width=12.cm]{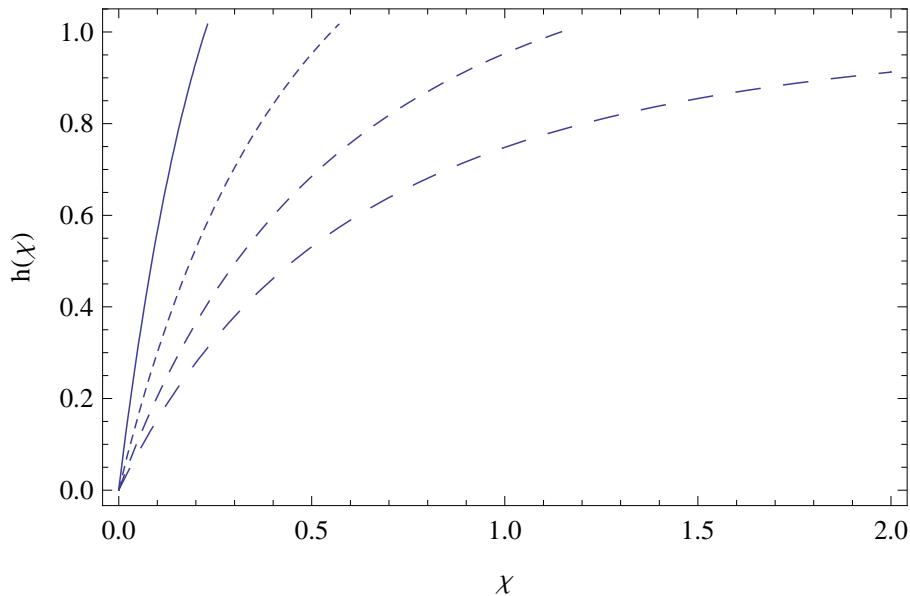}
\caption{Time evolution of the hadron fraction $h$ during the quark-hadron phase transition for $n=-0.74$ and different values of $\varepsilon$: $\varepsilon=-1/4$ (solid curve), $\varepsilon=-1/2$ (dashed curve), $\varepsilon=-3/4$ (long-dashed curve) and  $\varepsilon=-1$ (very long-dashed curve), respectively.}
\label{fig3}
\end{figure}

The variation of the hadron fraction given by Eq.~(\ref{eq:ht}), as a function of the dimensionless time parameter $\chi =\sqrt{\rho _Q}t_{Pl}t$ is represented, for different values of the parameter $\varepsilon$ and for $n$ fixed in Fig.~\ref{fig3}. The hadron fraction apparently gives an estimation for hadrons formed inside QGP. Having the expressions of $h(t)$ and $\dot{h}(t)$, the analytical forms for both $H$ and $a$ can be directly obtained. It is straightforward to show that the Hubble parameter during the phase transition can be expressed as
\begin{equation}
H=\frac{1}{\sqrt{3}m_{Pl}}\sqrt{\left[1+nh(t)\right]\rho _Q}.
\end{equation}

\begin{figure}[htb!]
\includegraphics[width=12.cm]{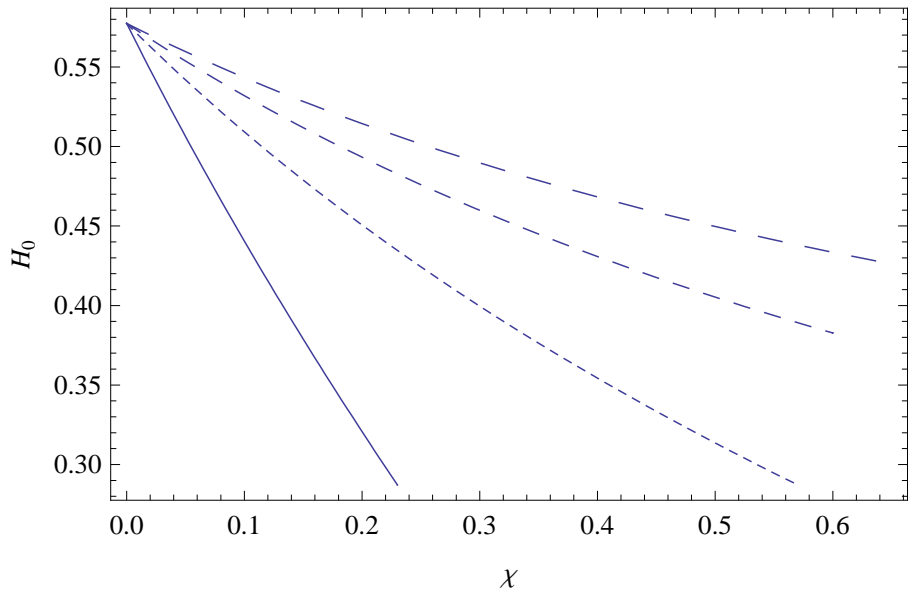}
\caption{Time evolution of the dimensionless Hubble parameter $H_0=H\times m_{Pl}/\sqrt{\rho _Q}$ during the quark-hadron phase transition as a function of the dimensionless time parameter $\chi =\sqrt{\rho _Q}t_{Pl}t$ for $n=-0.74$ and different values of $\varepsilon $: $\varepsilon=-1/4$ (solid curve), $\varepsilon=-1/2$ (dashed curve), $\varepsilon=-3/4$ (long-dashed curve) and  $\varepsilon=-1$ (very long-dashed curve), respectively.}
\label{fig3a}
\end{figure}

The variation of the dimensionless Hubble parameter $H_0=H\, m_{Pl}/\sqrt{\rho _Q}$ is represented, as a function of the dimensionless parameter $\chi =\sqrt{\rho _Q}t_{Pl}t$, for different values of the parameter $\varepsilon $ and for $n$ fixed in Fig.~\ref{fig3a}.

\subsection{Cosmological evolution in the hadronic phase (post quark-hadron phase transition)}

\begin{figure}[htb!]
\includegraphics[width=12.cm]{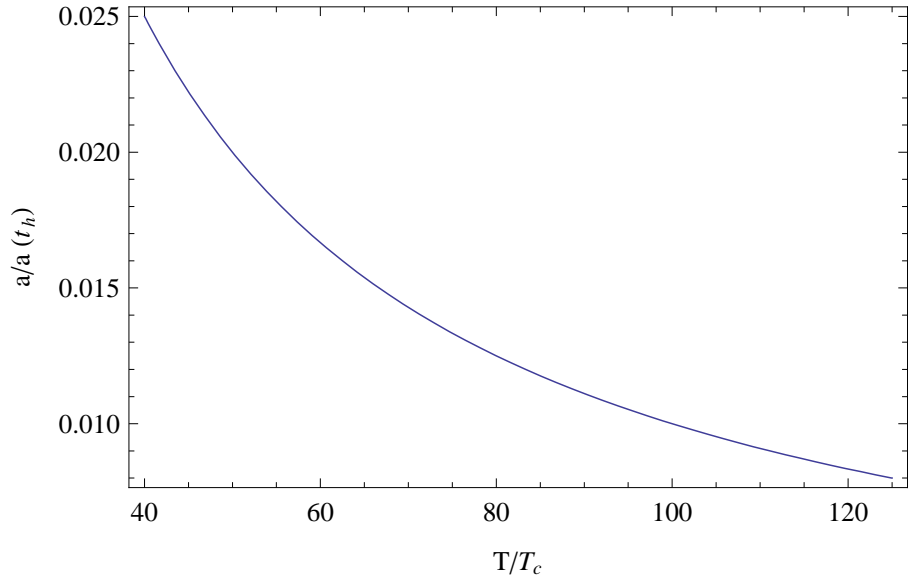}
\caption{The $T$ dependence of the scale factor $a$ during the hadronic phase. }
\label{fig3bb}
\end{figure}

\begin{figure}[htb!]
\includegraphics[width=12.cm]{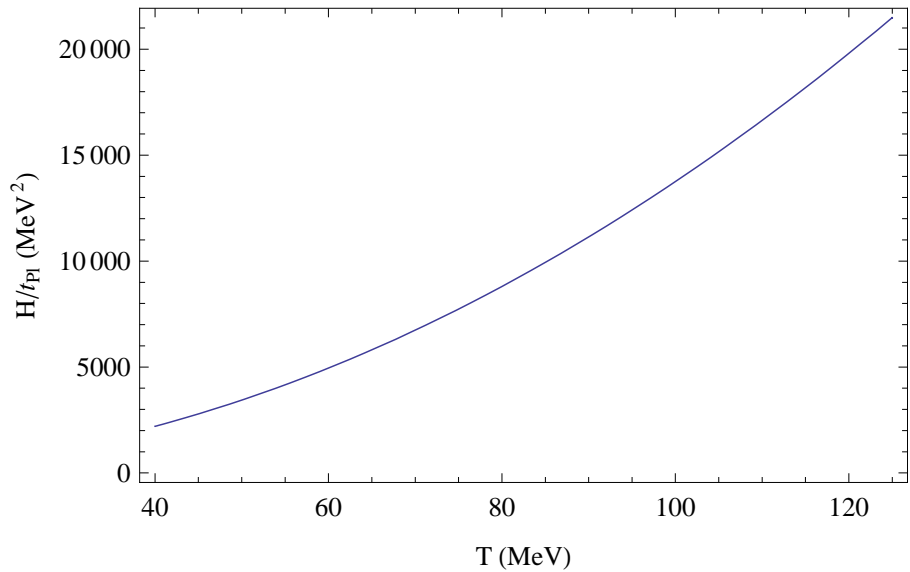}
\caption{The $T$-dependence of the Hubble parameter $H$ during the hadronic phase.}
\label{fig3bc}
\end{figure}

Finally, after the phase transition, the energy density of the pure hadronic matter is $\rho _{h}=3p_{h}=3a_{\pi }T^{4}$. The Bianchi identity Eq.~(\ref {drho2}) gives
\begin{equation} \label{eq:ah1}
a(T)=a\left(t_{h}\right)\; \frac{T_{c}}{T}.
\end{equation}
The time evolution of the temperature in the hadronic phase is governed by the equation
\begin{equation} \label{eq:Tdoth1}
\frac{dT}{dt}=-\frac{T}{\sqrt{3m_{Pl}}}\left(3\, a_{\pi
}\;  T^{4}\right)^{1/2}=-\frac{1}{m_{Pl}}\sqrt{a_{\pi }}\; T^3,
\end{equation}
giving a comoving  time
\begin{equation}
t-t_h =\frac{m_{Pl}}{2\,\sqrt{a_{\pi}}} \left(\frac{1}{T^2}-\frac{1}{T_c^2}\right).
\end{equation}
From Eq. (\ref{eq:ah1}) and (\ref{eq:Tdoth1}), the Hubble parameter reads
\begin{eqnarray}
H(T) &=& \frac{\sqrt{a_{\pi}}}{m_{Pl}}\; T^2, \\
H(t) &=& \frac{\sqrt{a_{\pi}}}{m_{Pl}}\frac{1}{T_c^{-2} + \left(2\sqrt{a_{\pi}}/m_{Pl}\right) (t-t_h)}.
\end{eqnarray}
During the hadronic phase, the density of the Universe varies with the time as
\begin{equation}
\rho_h(t)=\frac{3a_{\pi }}{m_{Pl}^2}\frac{1}{\left[T_c^{-2} + \left(2\sqrt{a_{\pi}}/m_{Pl}\right) (t-t_h)\right]^2}.
\end{equation}
The temperature dependence of the scale factor $a$ of the Universe during the hadronic evolution phase is presented in Fig.~\ref{fig3bb}. The temperature dependence of the Hubble parameter $H$ is represented in Fig.~\ref{fig3bc}.

\begin{figure}[htb!]
\includegraphics[width=12.cm]{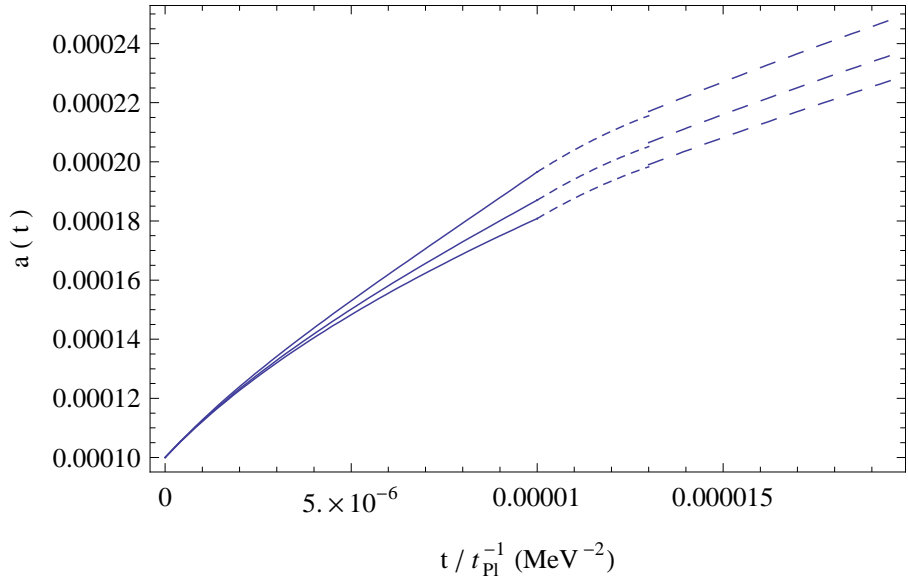}
\caption{Time evolution of the scale factor during the quark phase (solid line, quark-hadron phase transition (dashed line) and  hadron phase (long-dashed line), respectively, for several values of the bag constant $B$: $B^{1/4}=100$ MeV, $B^{1/4}=200$ MeV and  $B^{1/4}=250$ MeV. The numerical values of the scale factors raise with increasing $B$. 
The quark phase begins at $t=t_Q=0$, when the value of the scale factor is given as $a_0=a \left(t_Q\right)=a(0)=10^{-4}$.
The assumed critical transition temperature is $T_c=125$ MeV, the quark density is $\rho _Q=5\times 10^9$ MeV$^4$, while $\epsilon =-1/4$.}
\label{figfin}
\end{figure}

Finally, in Fig.~\ref{figfin} we present the time evolution of the scale factor $a$ of the Universe during the quark phase, the phase transition and the hadron phase, respectively, for several values of the bag constant $B$. We assume that the quark phase begins at a time $t=t_Q$, when the value of the scale factor of the Universe is $a=a\left(t_Q\right)$. The phase transition temperature is assumed to be $T_c=125$ MeV, with a corresponding quark matter energy density at the transition moment of $\rho_Q=5\times 10^9$ MeV$^4$. For the parameter $\varepsilon $ we have taken a value of $\varepsilon =-1/4$. As one can see from the figure, an increasing value of the bag constant accelerates, in the long term, the expansion of the universe.

\section{Phase transition in lattice QCD Simulations and heavy-ion collisions}
\label{qgpEoS}

Before introducing the QCD EoS, it is useful to study the similarities between heavy-ion collisions and the early Universe \cite{qgp-h5}. It is conjectured that the first-order phase transition, studied in section \ref{SecII}, might take place in the heavy-ion collisions and/or in lattice QCD simulations. Such a {\it prompt} transition seems to have fundamental astrophysical consequences. Its dynamics has been discussed in the previous section. Despite of the order of phase transitions, the QGP era seems not to be followed by an extreme expansion (inflation). This is apparently the case in heavy-ion collisions, because of the baryon number conservation and the limitation of baryon-to-photon ratio $(n_b-n_{\bar{b}})/n_{\gamma}\sim 10^{-11}$ \cite{wmap}. Therefore, $n_{\bar{p}}-n_p$ recently measured by ALICE experiment at $7\,$GeV can be used to estimate the photon number density, $n_{\gamma}\simeq 5.5\times10^{4}$, while in the CMB era, $n_{\gamma} \simeq 411.4 (T/2.73 \mathtt{K})\,$ cm$^{-3}$. Furthermore, the QGP era seems to be the last symmetry-breaking era of strongly interacting matter. By symmetry breaking, we mean deconfinement and chiral symmetry breaking and/or restoring, respectively.

In an isotropic and homogeneous background, the volume of the Universe is directly related to the scale factor $a(t)$, where $t$ is the comoving  time. Implementing a barotropic EoS for the background matter makes it possible to calculate - among others - the Hubble parameter $H(t)=\dot a(t)/a(t)$. Focusing the discussion on QCD era of the early universe, which likely turns to be fairly accessible in high-energy experiments, the equation of state is very well defined. In Ref. \cite{twfk7}, a viscous EoS for QGP matter has been introduced and different solutions for the evolution equation of $H$ have been worked out.
The ratio of baryon density asymmetry  to photon density, $\eta$, has been measured in WMAP data \cite{wmap}.
Then $n_{\bar{p}}-n_p$ from the ALICE experiment at $7\,$GeV can be used in order to estimate the photon number density, $n_{\gamma}\simeq 5.5\times10^{4}$, while in the CMB era, $n_{\gamma} \simeq 411.4 (T/2.73 \mathtt{K})\,$ cm$^{-3}$.

In a comoving  volume $V\sim a^3(t)$, the number density of noninteracting photons is supposed to remain constant. Therefore, $n_{\gamma}\sim 1/a^3(t)$. Nevertheless, when the Universe was expanding, $T$ decreases and $a^3(t)\,n_{\gamma}$ has to be affected. The previous values of $n_{\gamma}$ support this conclusion. There is a conserved quantity accompanying such a transition, namely, the entropy density $s$. In a {\it perfectly} closed system like the universe, $s$ likely remains unchanged.  From the first-law of thermodynamics \cite{twfk7} one can show that at vanishing chemical potential,
\bea
s(T) &=& \frac{P(T)+\rho(T)}{T} = \frac{a_1+1}{a_2}\;\rho(T)^{1-{a_3}},
\eea
implying that $s(T)$ is related to $\rho(T)^{1-{a_3}}$, where $a_1=0.319$, $a_2=0.718\pm 0.054$ and $a_3=0.23\pm 0.196$. This relation is valid at low energy, where the dominant degrees of freedom are given by hadron resonances. Baryon and boson relative abundances $(n(T)-\bar{n}(T))/(n(T)+\bar{n}(T))$ can be studied in hadron resonance gas (HRG) model. It is found that the abundance approaches $10^{-3}$. For instance, the kaon relative abundance is by about one order of magnitude higher than that of the proton. It is obvious that the abundances of the light elements ($^7$Li, $^4$He, $^3$He and $^2$H) produced in the early Universe are sensitive indicators of number density \cite{lightel}. Recent lattice QCD calculations~\cite{karsch07,Cheng:2007jq} give estimations for  the equation of state, temperature and bulk viscosity of hadronic and partonic matter at high temperatures.

%

As we will show below, the gravitational cosmological field equations, Eqs.~ (\ref{dH}) and (\ref{drho2}), relate the cosmological parameters, like the Hubble parameter $H$ and the scale factor $a$, to the energy density $\rho$. Again, the barotropic equations of state for the thermodynamic parameters are standard in analyzing the viscous cosmological models, whereas the equation for $\tau$ is a simple procedure to ensure that the speed of viscous pulses does not exceed the speed of light.

Using this set of equations seems to define the validity of this treatment. Apparently, it depends on the validity of the equations of states, which we have derived from the lattice QCD simulations at temperatures larger than $T_c\simeq 0.19~$GeV. Below $T_c$, as the Universe is cooled down, not only the degrees of freedom suddenly increase~\cite{Tawfik03}, but also the equations of state turn out to be the ones characterizing the hadronic matter. Such a phase transition - from QGP to hadronic matter - would characterize one end of the validity of our treatment. The other limitation is provided by the very high temperatures (energies), at which the strong coupling $\alpha_s$ entirely vanishes.

The lattice QCD simulations benefit from the rapid progresses achieved in computational facilities and algorithms. The accuracy of recent lattice results is comparable with the laboratory experiments. Currently, it is possible to perform lattice QCD simulations at almost physical quark masses. Recent results on QCD equations of states have been  reported in~\cite{bazazev}. It is apparent, that the influence of radiation and leptons on phase transition are minimized \cite{bono84}.


\subsection{Equations of state of the viscous quark-gluon plasma}

\begin{figure}[htb!]
\includegraphics[width=8.cm,angle=-90]{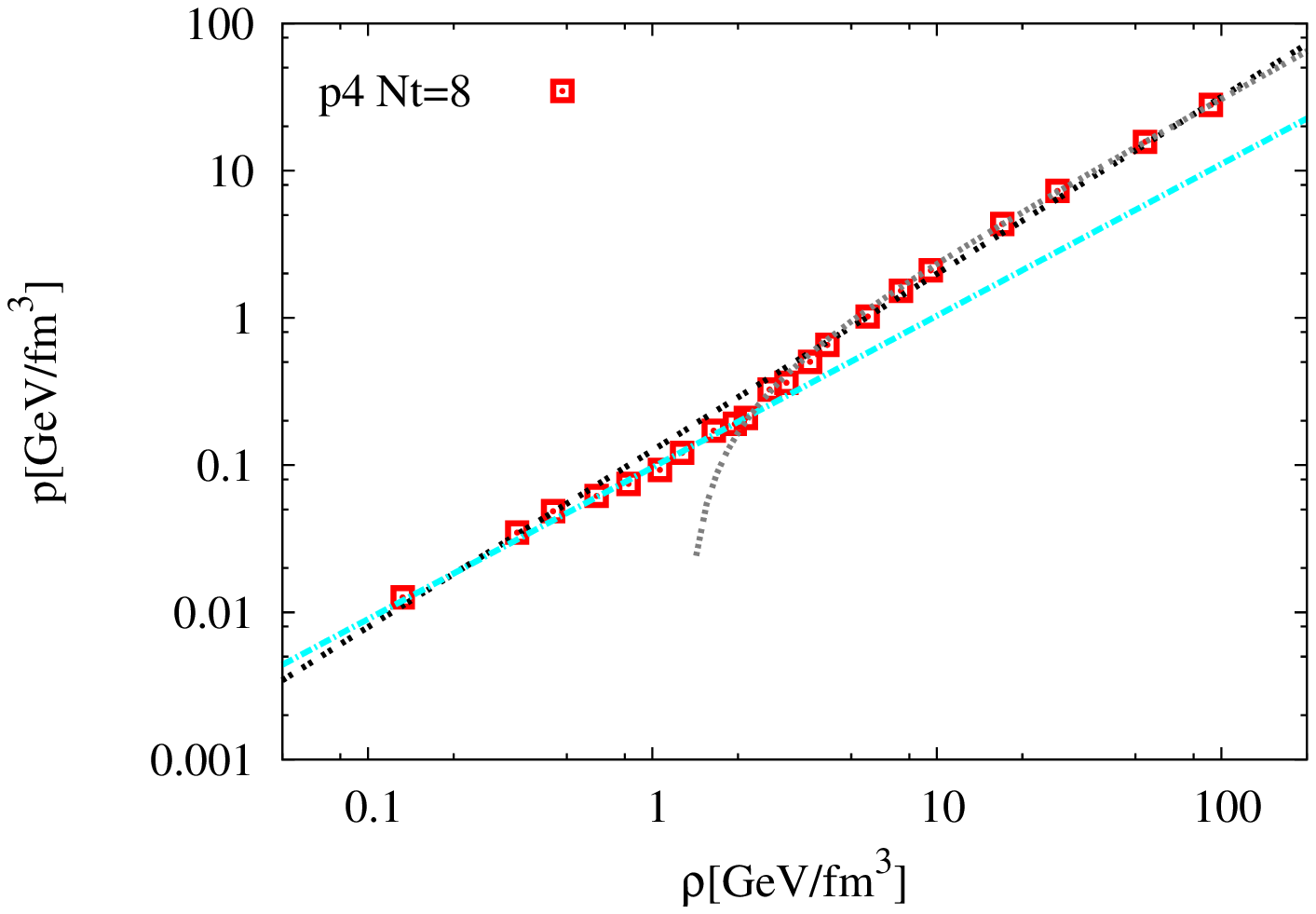}
\caption{The pressure density, $p$, is drawn in dependence on the energy density $\rho$. Both quantities are given in physical units. Symbols are lattice QCD calculations using $p4$ action and temporal lattice size, $N_{\tau}=8$ \cite{bazazev}. The dotted curve is the fitting in the quark phase, Eq. (\ref{aeq:p2}). The dash-dotted curve gives the fitting in the hadronic phase. The overall fitting is given by Eq. (\ref{aeq:p1}) (double-dotted curve). The small dip at $T_c$ seems to reflect the {\it slow} phase transition known as {\it crossover }.}
\label{afig1a}
\end{figure}

In this section, we give a list of barotropic equations of states deduced from the lattice QCD simulations \cite{bazazev} (an analytic crossover  phase transition is obtained) and the quasiparticle model \cite{qm1}. The latter is utilized when no lattice QCD results are available. Figure \ref{afig1a} depicts the pressure $p$ in dependence on the energy density $\rho$ in a wide range of temperatures, $1/2 <T/T_c<3$. Details on the lattice configurations are described in~\cite{bazazev}. It is obvious that the (barotropic) pressure - energy density  dependence  is almost linear referring to the nature of the phase transition from hadrons to quarks and vice versa. This confinement-deconfimement phase transition seems to be smooth i.e., simply continuous and takes place very slowly. This kind of transitions is a very moderate than the second-order one. The nature of the phase diagram in lattice QCD has been discussed in~\cite{taw05}.
In Fig. \ref{afig1a}, the dashed line represents the fitting in the entire $T$-region. Ignoring the dip around $T_c$, the results can be fitted as a power law,
\begin{equation} \label{aeq:p1}
p(\rho) = \alpha_1 \rho^{\alpha_2},
\end{equation}
where $\alpha_1=0.178\pm 0.009$ and $\alpha_2=1.119\pm 0.011$, respectively. In the hadronic phase i.e., at temperatures $<T_c$, the previous power law dependence seems to remain valid. Little changes appear in the parameters; $\alpha_1=0.096\pm 0.003$ and $\alpha_2=1.03\pm0.04$. In the quark phase i.e., at temperatures $>T_c$, the following polynomial
\begin{equation} \label{aeq:p2}
p(\rho)=-\frac{1}{3}+\alpha_1 \rho^{\alpha_2},
\end{equation}
describes this barotropic equation of state, where $\alpha_1=0.221\pm 0.004$ and $\alpha_2=1.072\pm 0.005$.

Apparently, expressions (\ref{aeq:p1}) and (\ref{aeq:p2}) imply that the speed of sound drastically changes with the changes in the phases:
\begin{itemize}
\item hadron and quark phase: $c_s^2=\partial p/\partial \rho \simeq 0.199\; \rho^{0.119}$ i.e., the sin the peed of sound depends on the energy density. The latter has a nonmonotonic behavior when going from hadronic to partonic phases and vice versa.  
\item in the hadron phase: $c_s^2\simeq 0.098$,
\item in the quark phase: $c_s^2\simeq 0.237$.
\end{itemize}

\begin{figure}[htb!]
\includegraphics[width=8.cm,angle=-90]{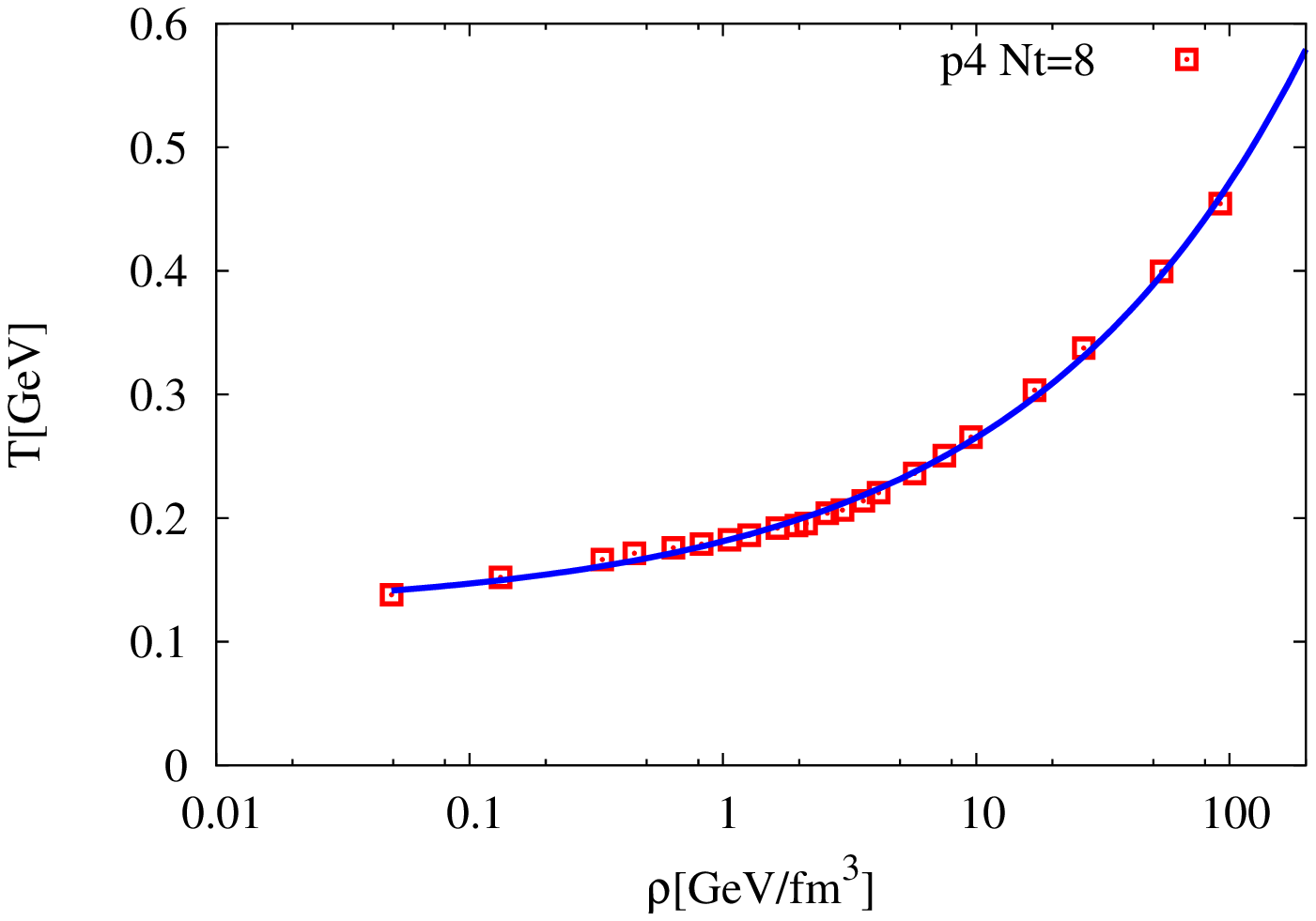}
\caption{The symbols are the lattice QCD calculations for the barotropic dependence of temperature $T(\rho)$ using $p4$ action and temporal lattice size, $N_{\tau}=8$. Details on the lattice configurations are given in Ref. \cite{bazazev}.
The solid curve gives the fitting according to Eq. (\ref{aeq:t}). }
\label{afig:1b}
\end{figure}

Figure~\ref{afig:1b} presents the barotropic dependence of $T$ as calculated in lattice QCD. The relation can nicely be fitted by the polynomial
\begin{equation} \label{aeq:t}
T(\rho)=\beta_1+\beta_2\rho^{\beta_3},
\end{equation}
where $\beta_1=0.123\mp 0.004$, $\beta_2 = 0.058 \pm 0.0038$ and $\beta_3 = 0.39 \pm 0.013$. Again, the specific heat in the whole phase of hadrons and quarks seems to depend on the energy density, strongly 
\bea
c_V &=& \frac{\partial \rho}{\partial T} \sim  44.247\; \rho^{0.61}.
\eea
In determining this value, the volume $V$ is conjectured to remain unchanged.

\begin{figure}[htb!]
\includegraphics[width=8.cm,angle=-90]{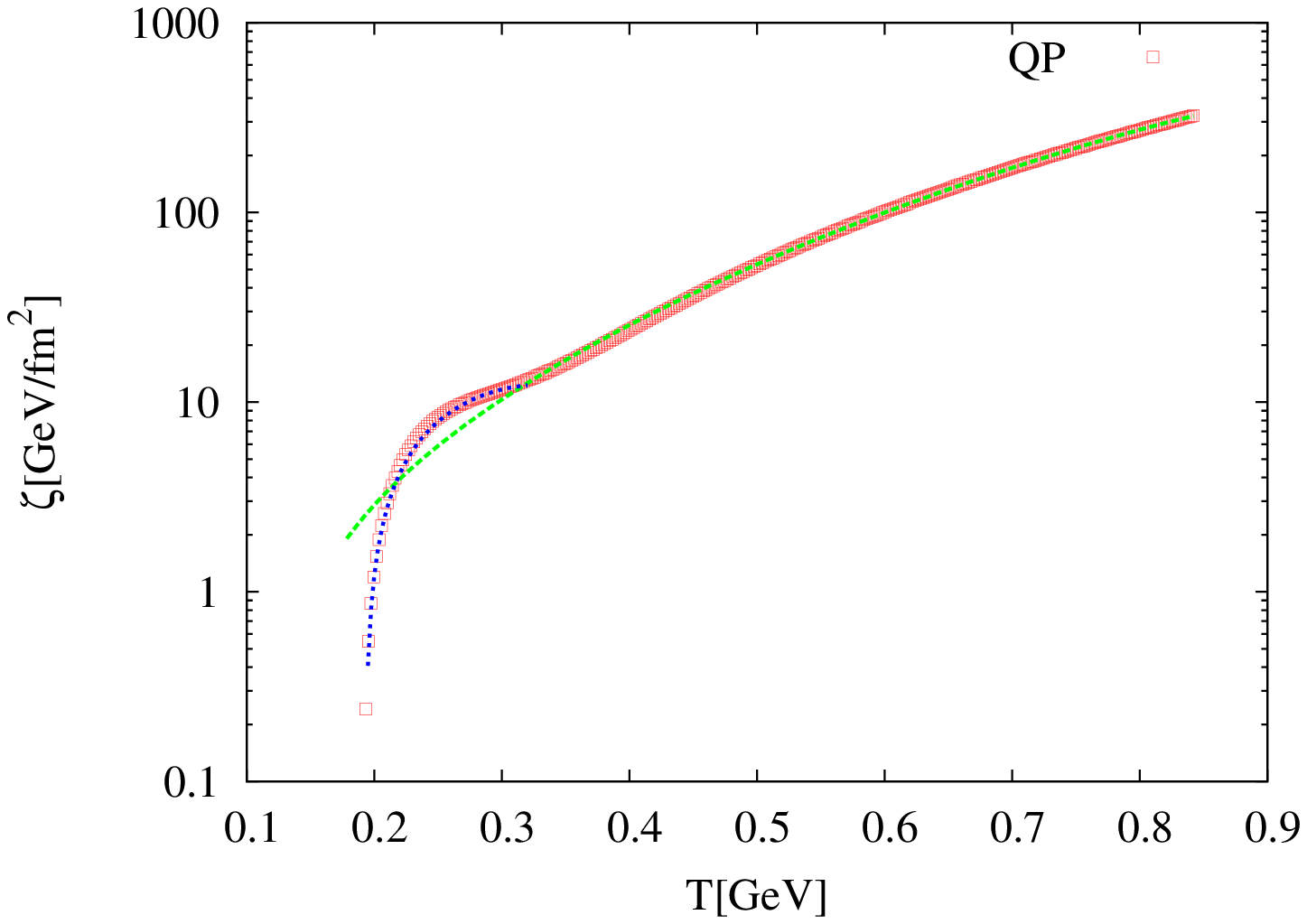}\\
\includegraphics[width=8.cm,angle=-90]{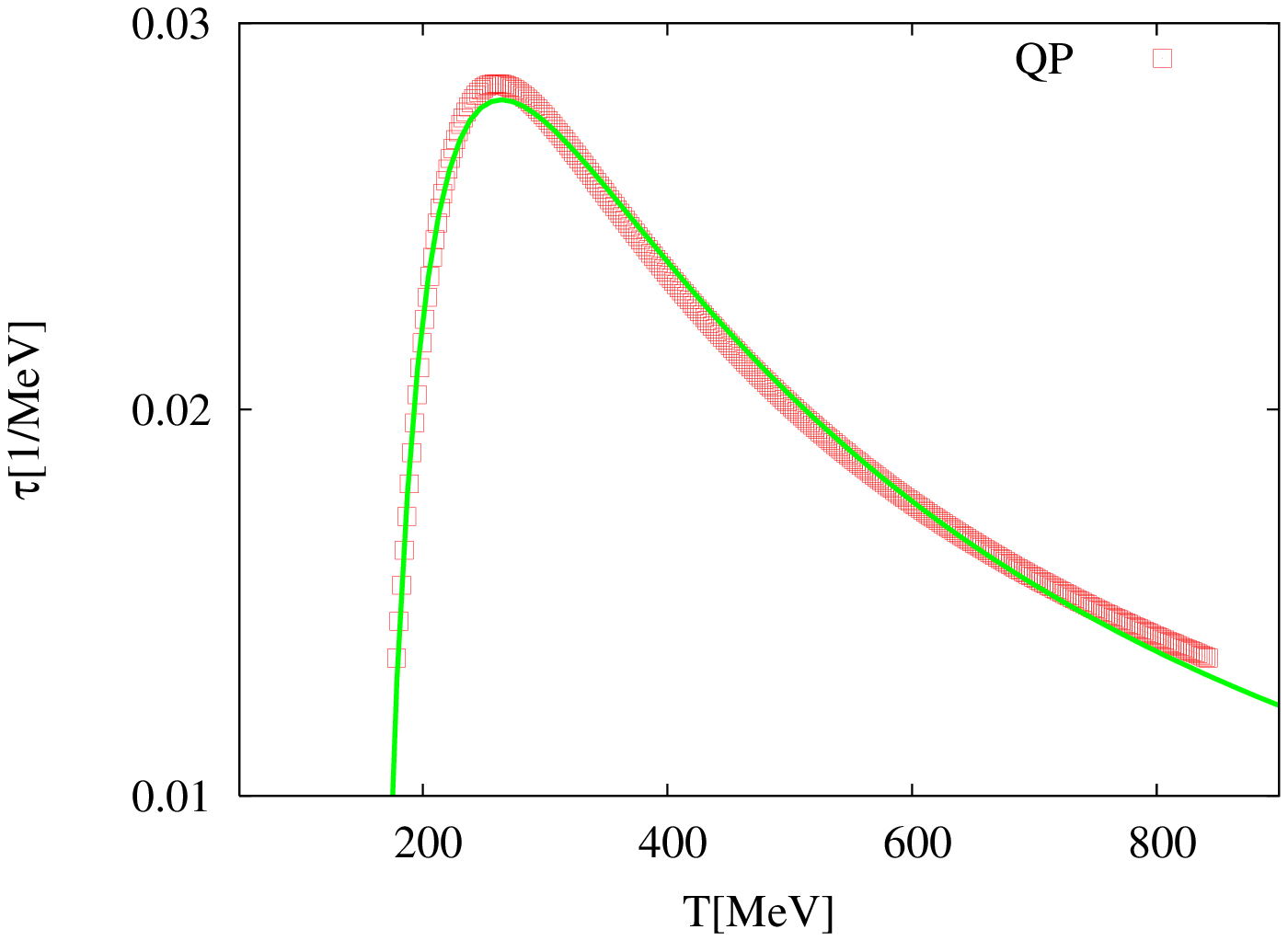}
\caption{The top panel depicts the dependence of the bulk viscosity $\zeta(T)$ on the temperature $T$ as deduced from the quasiparticle model (symbols) \cite{qm1,redlich:qm}. The dotted curve gives the fitting of results related to the hadronic phase. The fitting of the results in the partonic phase is given by the dashed curve, Eq. (\ref{aeq:zeta2}). The two regions meet at  $T_c\sim 270~$MeV. Amazingly, this value finds its root in lattice QCD. The quenched lattice QCD calculations, where the quark masses are supposed to be very heavy, predict that the critical temperature separating hadrons from QGP has the same value.  In bottom panel, relaxation times $\tau$ is drawn against $T$ in MeV units. The solid curve gives the fitting according to Eq. (\ref{aeq:tau2}). Again, there are two separate regions. In the first one, $\tau$ raises with increasing $T$. While, in the second region, $\tau$ decreases with increasing $T$. Such a nonmonotonic behavior is characterized at $T_c\sim 270~$MeV. 
}
\label{afig:2}
\end{figure}

Based on the quasiparticle model~\cite{redlich:qm}, which is an effective model used to reproduce the lattice QCD results on various thermodynamic and transport properties~\cite{kampfer}, the bulk viscosity $\xi$ reads
\begin{eqnarray}\label{aeq:zeta}
\xi (T) &=& \frac{g}{2\pi^2} \frac{1}{3T}\int_0^{\infty} \vec{p}^2dp \frac{\tau}{\epsilon} f_0(1+f_0) 
\left[\frac{\vec{p}^2}{3\epsilon} - \left(\epsilon - T \frac{\partial \epsilon}{\partial T}\right)\frac{\partial p}{\partial \rho}\right]
\left\{2T^2\frac{\partial\Pi(T)}{\partial T^2}-\Pi(T)\right\},
\end{eqnarray}
where $g$ is the degeneracy factor of quarks, gluons and their antiparticles. The function $f_0$ gives the distribution function $(\exp(\epsilon)\pm 1)^{-1}$ of boson and fermion particles, respectively. The quantity $\epsilon=[\vec{p}^2+\Pi(T)]^{1/2}$ stands for the effective dispersion relation of single particles.  It depends on the particle mass $\Pi(T)$ which in turn varies with the effective coupling $G(T)$. Therefore, the effective coupling $G(T)$ plays an essential role in this model. It has to be adjusted to reproduce lattice QCD results. In top panel of Fig.~\ref{afig:2}, the bulk viscosity coefficient $\xi$ is given as a function of $T$.
The results are fitted very well as  
\begin{equation} \label{aeq:zeta2}
\xi(T) = \ln\left(\frac{T-\gamma_1}{\gamma_2}\right) \left[\gamma_3 + \gamma_4 \left(T-\gamma_5\right)^{\gamma_6}\right].
\end{equation}
At high $T$, the fitting parameters are $\gamma_1=1.042 \pm0.067$, $\gamma_2=-0.0035 \pm  0.0059$, $\gamma_3=-0.329\pm  0.058$, $\gamma_4=25.666\pm 1.521$ and $\gamma_5 = -0.367 \pm 0.0159$. At low $T$, the fitting parameters read $\gamma_1= 0.801\pm0.595$, $\gamma_2= -0.352 \pm 0.558$, $\gamma_3= -0.350 \pm 0.174$, $\gamma_4=303.582 \pm 3.126$ and $\gamma_5= 0.189 \pm 0.054$. In this region the exponent $\gamma_6=1$. In the high-$T$ region, $\gamma_6=6$.

Again, in quasiparticle model \cite{redlich:qm,qm1}, the relaxation time reads
\begin{equation}\label{aeq:tau}
\tau^{-1}(T) = \frac{a_{\zeta}}{32 \pi^2} T G^4(T) \ln\left(\frac{a_{\zeta}\pi}{G^2(T)}\right),
\end{equation}
where $a_{\zeta}=6.8$~\cite{redlich:qm}. The results are drawn in bottom panel of Fig. \ref{afig:2}. Fitting of these results leads to  
\begin{equation} \label{aeq:tau2}
\tau(T) =\delta_1 \ln \left(-\frac{\delta_2 T-\delta_3}{\delta_4}\right)^{\delta_5}  \frac{\delta_6}{T^{\delta_7}},
\end{equation}
where $\delta_1=2.362 \pm1.318$, $\delta_2= -0.022\pm 0.056$, $\delta_3=-3.176\pm 1.05$, $\delta_4=0.435 \pm0.126$, $\delta_5=2.362\pm0.318$, $\delta_6=3$ and $\delta_7=1.25\pm 0.12$.

\begin{figure}[htb!]
\includegraphics[width=8.cm,angle=-90]{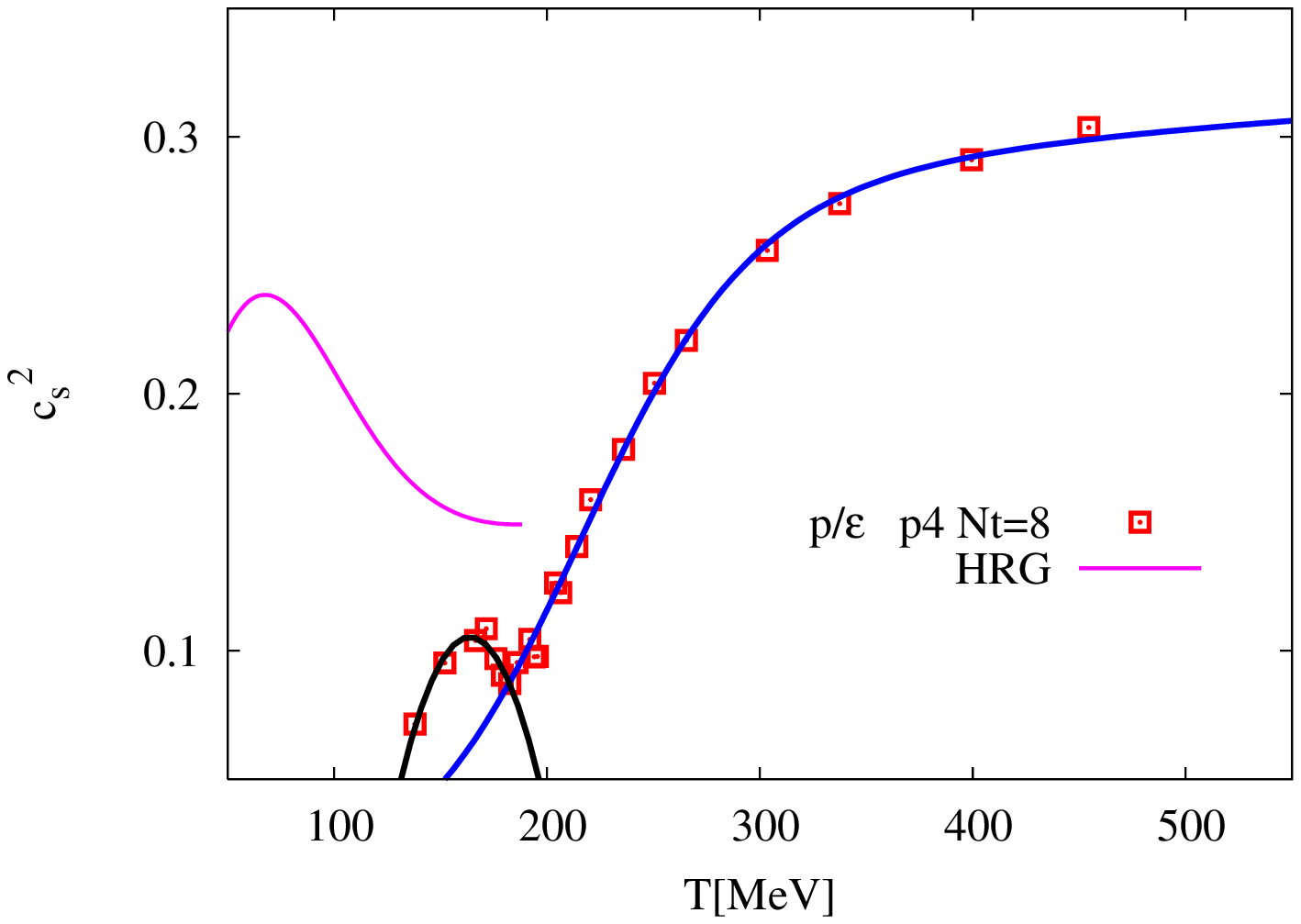}
\caption{The speed of sound $c_s^2={\partial p/\partial \rho}$ is drawn versus $T$ in the physical units. The symbols represent the results from lattice QCD calculations~\cite{bazazev}, using $p4$ action and temporal size $N_{\tau}=8$. Below $T_c$, results from the hadron resonance gas model are given (solid curve on the left). The other two curves are fittings for lattice QCD data.}
\label{afig:1c}
\end{figure}

From the two expressions (\ref{aeq:zeta2}) and (\ref{aeq:tau2}), it is obvious that the barotropic relations of $\xi$ and $\tau $ are related to each other. Such a relation has been modeled by the projection operator method \cite{dirk2010} as
\bea \label{eq:tauRischke}
\tau&=&\xi\, \frac{T}{\left(\frac{1}{3}-c_s^2\right) -\frac{2}{9} (\rho-3p)}.
\eea
Furthermore, the bulk stress is to be related to the distribution function of relaxation time \cite{andrew}. Such a dependence has to follow the causality principle and fits perfectly with the laws of thermodynamics \cite{maartens}. The speed of sound $c_s^2=\partial p/\partial \rho$, can be taken from the lattice QCD simulations~\cite{bazazev}. The results of $c_s^2(T)$ are given in Fig.~\ref{afig:1c}. Below $T_c$, the lattice results show a small peak. Remarkable work has been devoted to accurate its location and altitude. The results from the hadron resonance gas model are given, as well. Although the appearance of the peak, the disagreement is not to be neglected.  With reference to the {\it restricted} causality principle, the nonmonotonic behavior of $c_s^2$ below $T_c$ would be explained in the light of: 
\begin{itemize}
\item baryon and strange degrees of freedom which would play an essential role in reproducing $c_s^2(T,\mu_b)$, where $\mu_b$ is the baryo-chemical potential,
\item the interpolation of both entropy $s(T,\mu_b)$ and specific heat $c_V(T,\mu_b)$ which has been suggested to partly explain nonmonotonic behavior below $T_c$, 
\item the condition(s) deriving the chemical and thermal freeze-out which would enlighten such a behaviour, 
\item the interactions between the constituents of the hadronic phase are conjectured which would be able to explain the nonmonotonic entropy and specific heat production and 
\item the time-varying equation of state in the hadronic phase which refers to out-of-equilibrium processes, while their modification in thermal and dense matter would refer to symmetry changes.
\end{itemize}


\subsection{Bulk Viscosity in the Hadronic Phase}

The treatment of bulk viscosity in Hagedorn fluid has been studied in Ref. \cite{twfk6}. Such a fluid is conjectured to be composed of hadrons and resonances with masses $m<2\,$GeV. The treatment is based on the relativistic kinetic theory formulated under the relaxation time approximation. The {\it in-medium} thermal effects on bulk (and shear) viscosity and the averaged relaxation time with and without the excluded-volume approach are deduced. It has been suggested that the dynamics of the heavy-ion collisions, the nonequilibrium thermodynamics and the cosmological models, require thermo- and hydrodynamic equation(s) of state. When assuming vanishing chemical potential and the heat conductivity, the bulk viscosity in thermal medium reads
\begin{eqnarray}
\xi(T) &=& \frac{g}{2\pi^2} \frac{\tau}{T} \sum_i\rho(m_i)\int_{0}^{\infty} n_0(1+n_0)\left(c_s^2 \varepsilon_i^2 - \frac{1}{3}\vec{p}\, ^2 \right)^2 p^2 dp,
\end{eqnarray}
where $\rho(m_i)$ is the Hagedorn mass spectrum $\rho(m)$, which implies growth of the hadron mass spectrum with increasing the resonance masses.
\begin{eqnarray}\label{eq:rhom}
\rho(m) &=& A\; \left(m_0^2+m^2\right)^{k/4} \exp(m/T_H),
\end{eqnarray}
with $k=-5$, $A=0.5\,$GeV$^{3/2}$, $m_0=0.5\,$GeV and $T_H=0.195\,$GeV. The number density
$n_0$ is related to the deviation of energy-momentum tensor from its local equilibrium $\delta T^{\mu\nu}$. Such a deviation is corresponding to the difference between the distribution function near and at equilibrium, $\delta n=n-n_0$. The latter can be determined by the relaxation time approximation with vanishing external and self-consistent forces \cite{V-hadron2,relaxx}
\begin{eqnarray}
\delta n(p,T) &=& - \tau(T)\, \frac{p^{\mu} }{\vec{p}\cdot\vec{u}} \partial_{\mu} n_0(p,T).
\end{eqnarray}
The nonequilibrium number density $n(p,T)$ is to be decomposed using the relaxation time approach into $n=n_0+\tau n_1+\cdots$. Alternatively, as $n(p,T)$ embeds the 1$^{st}$-rank tensor $u$, $\delta T^{\mu\nu}$ can be decomposed into $u$ \cite{V-hadron2} in order to deduce its spatial components.

The relaxation time depends on the relative cross section as
\begin{eqnarray}
\tau(T) &=& \frac{1}{n_f(T)\langle v(T)\sigma(T)\rangle},
\end{eqnarray}
where $v(T)$ and $n_f(T)$ is the relative velocity of two particles in case of binary collision and the density of each of the two species, respectively. The thermal-averaged transport rate or cross section is $\langle v(T)\sigma(T)\rangle$.

The ratio of bulk to shear viscosity, $\xi/\eta$, can be related to the speed of sound $c_s^2$ in a gas composed of massless pions. Apparently, there are essential differences between this system and the Hagedorn fluid. According to \cite{etaxiratio1}, the ratio of $\xi/\eta$ in $N=2^*$ plasma is conjectured to remain finite across the second-order phase transition. In Hagedorn fluid, the system is assumed to be drifted away from equilibrium and it should relax after a characteristic time $\tau$. Should we implement a phase transition in Hagedorn fluid, then $\tau\propto \xi^{z}$, where $z$ is the critical exponents, which likely diverges near $T_c$.

\subsection{Deconfinement and chiral phase transitions (crossover ) in lattice QCD simulations}

Remarkable advances have been made in studying the equilibrium properties of the phase transitions. Obviously, the phase transition is coupled with symmetry breaking and out-of-equilibrium. Therefore, it is natural to turn our attention to the consequences when the system is enforced to go through an out-of-equilibrium phase transition. Thermodynamically, the  first and second order phase transitions are described by continuous first and second derivative of the free energy, respectively. The {\it infinite} order phase transition is also continuous. But it breaks no symmetry. A famous example for it is the Kosterlitz-Thouless transition in the two-dimensional $XY$ model \cite{kt1}. The crossover phase transition of lattice QCD simulations is  likely a continuous one. An out-of-equilibrium state is reached when the system in deviated from its equilibrium state by applying an instantaneous perturbation. The system will relax to its equilibrium state by dissipating the energy transferred during the transition \cite{alberto}. Relating the amplitude of this dissipation to the amplitude of fluctuations in equilibrium dates back to Einstein's work on the Brownian motion in 1905. Lars Onsager established a hypothesis that the relaxation of a macroscopic nonequilibrium perturbation follows the same laws which govern the dynamics of fluctuations in equilibrium systems. In other words, the regression of microscopic thermal fluctuations at equilibrium follows the macroscopic law of relaxation of small nonequilibrium disturbances \cite{onsager}.

The lattice QCD simulations turn out to be an accurate tool to study - among others - the thermodynamics of the hadronic and partonic matter up to temperatures of couple $T_c$; the critical temperature $T_c$ ~\cite{Fodor:2001au, deForcrand:2002ci, Allton:2002zi,
  D'Elia:2002gd, Karsch:2000kv, Karsch:2001cy, Gavai:2003mf}. For two quark flavors ($n_f=2$) the phase transition is second-order or a rapid crossover . $T_c\simeq 173\pm8\;$MeV. For $n_f=3$, the phase transition is first-order  and $T_c\simeq 154\pm8\;$MeV. For $n_f=2+1$, the transition is again crossover and $T_c\simeq 173\pm8\;$MeV. For the pure gauge theory, $T_c\simeq 271\pm2\;$~MeV and the phase transition is first-order. In all these lattice QCD simulations, the quark masses are heavier than their physical values. At physical masses, the critical temperature for $n_f=2+1$ is $\simeq 200~$MeV. Apparently, we conclude that the order of the deconfinement phase transition can be either first, or second or crossover (infinite). It depends on the quark flavors and their masses. The extreme conditions in the early Universe likely affect the properties of the hadronic and partonic matter.

The chiral phase transition is assumed to accompany the deconfinement one, especially at vanishing chemical potential. It is expected that the restoration of the chiral symmetry breaking takes place in full-perturbative and nonperturbative QCD at high temperatures, if the matter is assumed to be exclusively built of light and strange quarks \cite{ioffe}. In perturbative QCD, the chiral symmetry is valid for massless quarks. It is entirely broken in the hadronic phase. It not yet completely clear what is the order of phase transition between hadronic and partonic QCD phases when the broken symmetry is restored at finite temperatures and densities. Different lattice QCD simulations, mainly referring to chiral condensate and chiral susceptibilities \cite{latticechiral} indicate that the chiral phase transition is of the second order at vanishing chemical potential:
\bea
\mathtt{SU}(n_f)_l \times \mathtt{SU}(n_f)_r \rightarrow \mathtt{SU}(n_f)_V.
\eea
The chiral condensate vanishes at the limit $m_q\rightarrow 0$ \cite{tawdom}. Below $T_c$, the chiral condensate entirely vanishes, as well. It is finite above $T_c$,
\bea
\langle \psi \bar{\psi}\rangle =-\frac{T}{V} \frac{\partial}{\partial m_q} \ln Z,
\eea
where $\ln Z$ is the partition function describing the system. The chiral perturbation theory proved to be a very important method in determining some essential observables in QCD, which are dominated at low temperature, such as the masses of pseudoscalar mesons, their decay constants and the chiral observables. It provides an explanation why pseudoscalar mesons are very light. The Goldstone theorem states that for each generator of a spontaneously broken symmetry, there exists a massless Goldstone boson $\phi$ with spin $0$ and  with symmetry properties that are related to those of the symmetry transformation. The Goldstone bosons of the chiral perturbation theory are just the pseudoscalar mesons. This can be utilized as a signature for the phase transition.

\section{Dynamics of the bulk viscous quark-gluon plasma filled universe}\label{cosm}

In the following we consider the cosmological evolution of the viscous quark-gluon plasma filling the Universe in the framework of both Eckart and the full causal approaches to dissipative processes.

\subsection{Evolution of the Hubble parameter in the Eckart model}

\begin{figure}[htb!]
\includegraphics[width=15.cm]{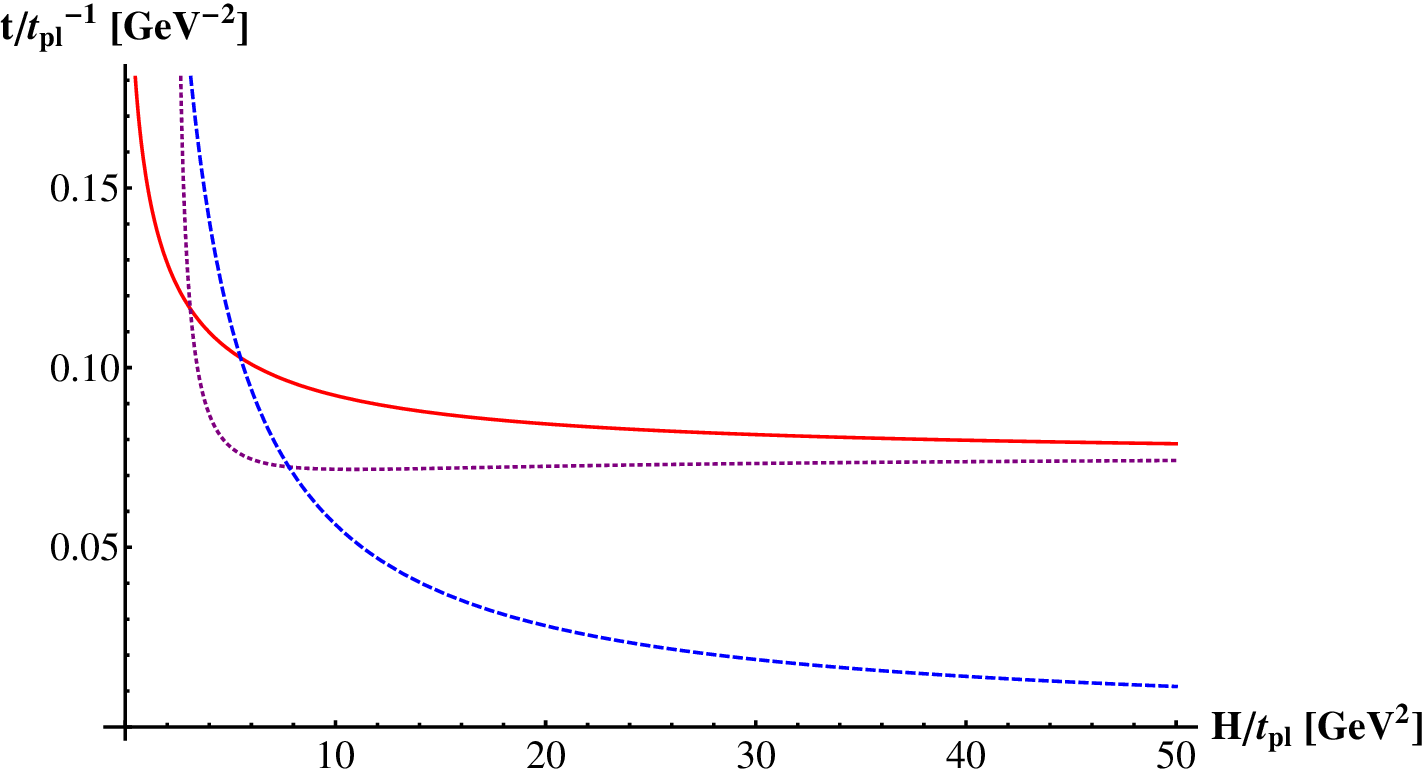}
\caption{The comoving  time $t$ is drawn (solid curve) in dependence on the Hubble parameter $H$, Eq. (\ref{eq:eckrt1}). The background fluid is characterized by Eckart theory. The dotted curve gives the results when viscous QGP equation of state is implemented \cite{twfk1,twfk2,twfk3,twfk4,twfk5,twfk8,twfk9}. The dashed curve shows the results when the background geometry is assumed to be filled with an ideal gas.  The Planck scale is given in physics units.
}
\label{afig:eckart1}
\end{figure}

Substituting the barotropic expressions (\ref{aeq:p1}), (\ref{aeq:t}) and (\ref{aeq:zeta2}) in Eq.~(\ref{eq:evlu}) and by assuming for the bulk viscous pressure the Eckart relation, given by Eq.~(\ref{entr}), we obtain for the evolution of the Hubble parameter the equation
\bea
\dot H + \frac{3}{2\,A}
\Bigg\{A\, H^2 + \alpha_1\, A^{\alpha_2}\, H^{2\alpha_2} -
 3 \ln\left[\frac{\beta_2\, A^{\beta_3}\, H^{2\beta_3}-\gamma_1}{\gamma_2}\right] 
\left(\gamma_3 + \gamma_4\left(\beta_2\, A^{\beta_3}\, H^{2\beta_3} - \gamma_5\right)^{\gamma_6}\right)\, H\, \Bigg\} &=& 0,
\eea
where $A=3/(8\pi G)$. As given in the introduction, the Planck units are given by this parameter, $A=1.778 \times 10^{37}\;$ GeV$^{2}$.
This differential equation can be solved, analytically, when assuming that
\begin{equation}
H^{2\beta_3}=(\gamma_1+\gamma_2)/\beta_2 A^{\beta_3}.
\end{equation}
Then, in terms of $H$, the comoving  time reads
\be
t = 
    2A \frac{\ln\left[-A \left(1+\alpha_2-3\beta_2\gamma_4\right)+3\left(\gamma_3-\gamma_4\gamma_5\right)/H\right] }{9 (\gamma_3 -\gamma_4 \gamma_5)}.
\label{eq:eckrt1}
\ee
Figure \ref{afig:eckart1} shows the dependence of $t$ on $H$. It describes a universe, where its background fluid is characterized by Eckart theory. The three curves differentiate between different types of the background matter. A discussion about the effect of background matter is given in section \ref{sec:mt}. The collisionfree and nonviscous background matter is given by the dashed curve. Solid and dotted curves describe the $t-H$ relation when the background geometry is filled with viscous hadron-QGP and nonviscous QGP, respectively. At small $H$ values, there are obvious differences between the latter types of matter and between them and the ideal matter. At large $H$ values, the comoving  time behaves very smooth with $H$, although hadron-QGP results in larger $t$ than QGP. In both of them, $t$ is larger than in ideal matter.

\subsection{Evolution of the Hubble parameter in the full causal approach}

Comparing Eq.~(\ref{eq:tauRischke}) with the expressions (\ref{aeq:zeta2}) and (\ref{aeq:tau2}) makes it quite apparent to have a barotropic expression for the relaxation time \cite{Ma95}
\bea \label{eg:tau}
\tau &=& \xi/\rho,
\eea
i.e., the relaxation coefficient for the transient bulk viscous effect is referred to as the relaxation time.

\begin{figure}[htb!]
\includegraphics[width=8.cm]{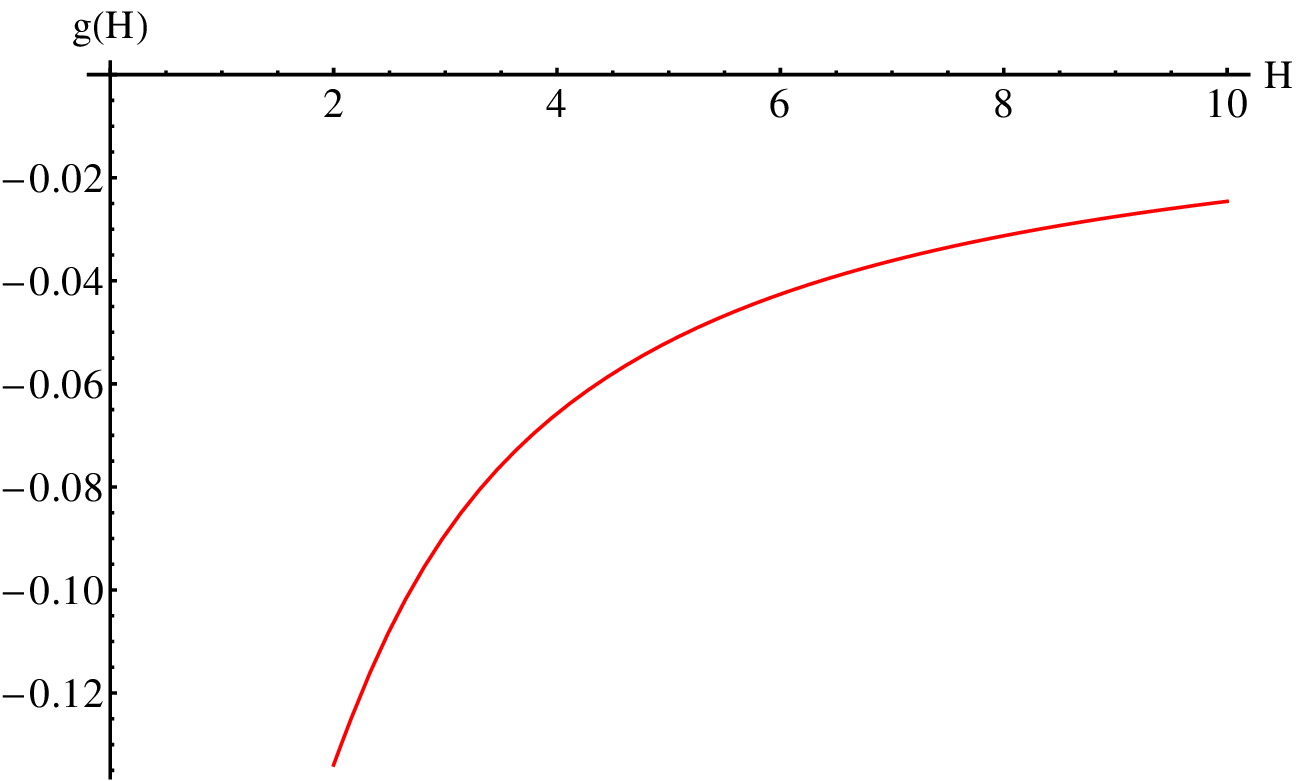}
\includegraphics[width=8.cm]{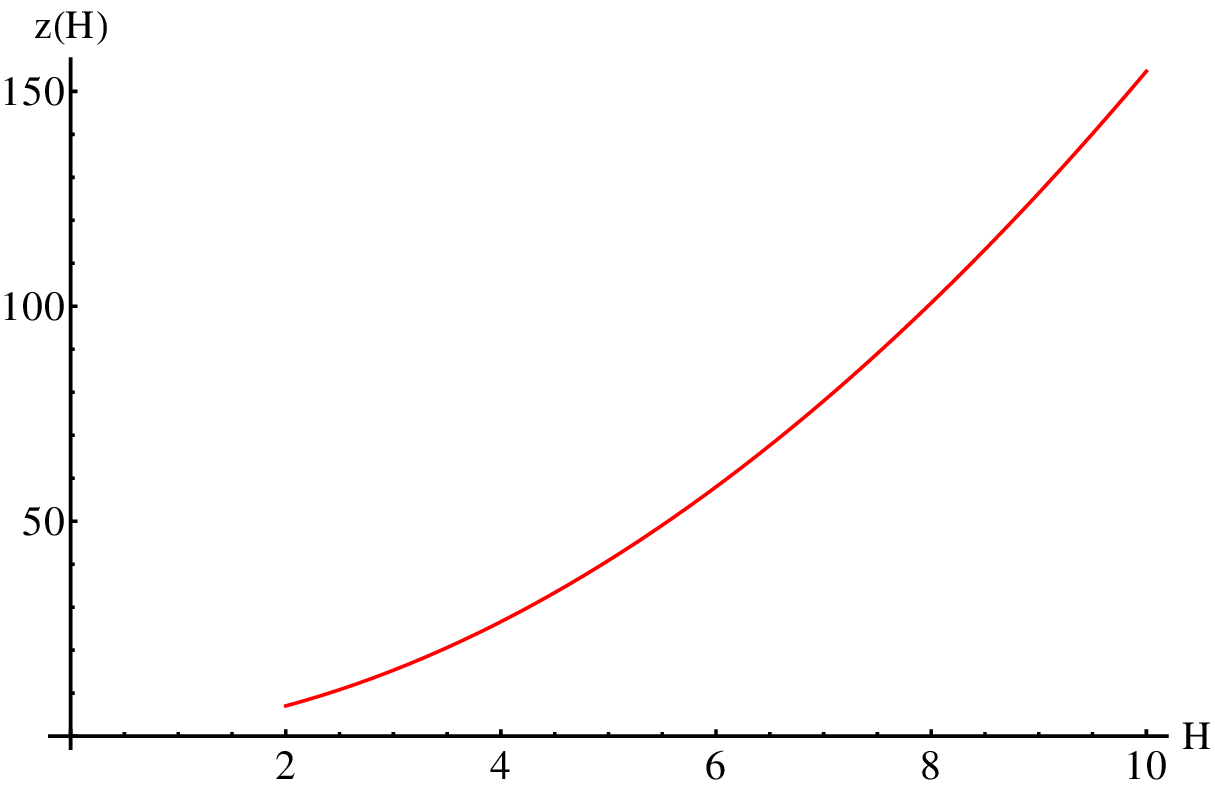}
\caption{Parametric functions given in Eqs. (\ref{eq:isgH}) and (\ref{eq:iszH}) are depicted in dependence on $H$. This illustration gives an indication about the dependence of the function $g$ on the independent functional parameter $z$. Such a dependence is given in Fig. (\ref{afig:is-gz1}).
}
\label{afig:is1}
\end{figure}

\begin{figure}[htb!]
\includegraphics[width=16.cm]{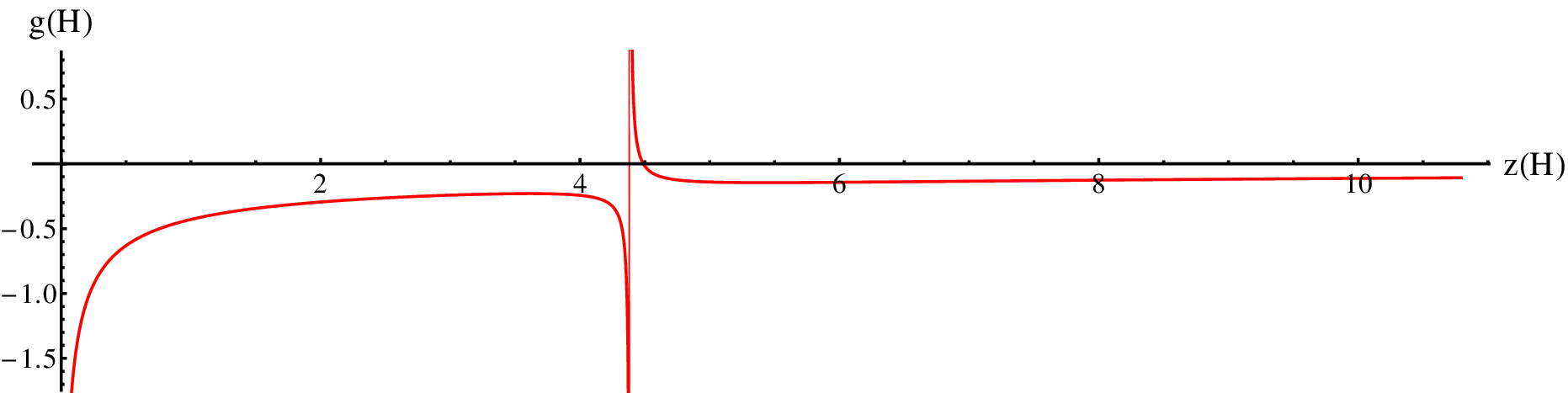}
\caption{The parametric dependence of the function $g$ on the independent functional parameter $z$. Apparently, it has a nonmonotonic behaviour. At $H$-values, where this study is valid, the dependence is apparently linear, $g(z)\propto z$.
}
\label{afig:is-gz1}
\end{figure}

\begin{figure}[htb!]
\includegraphics[width=15.cm]{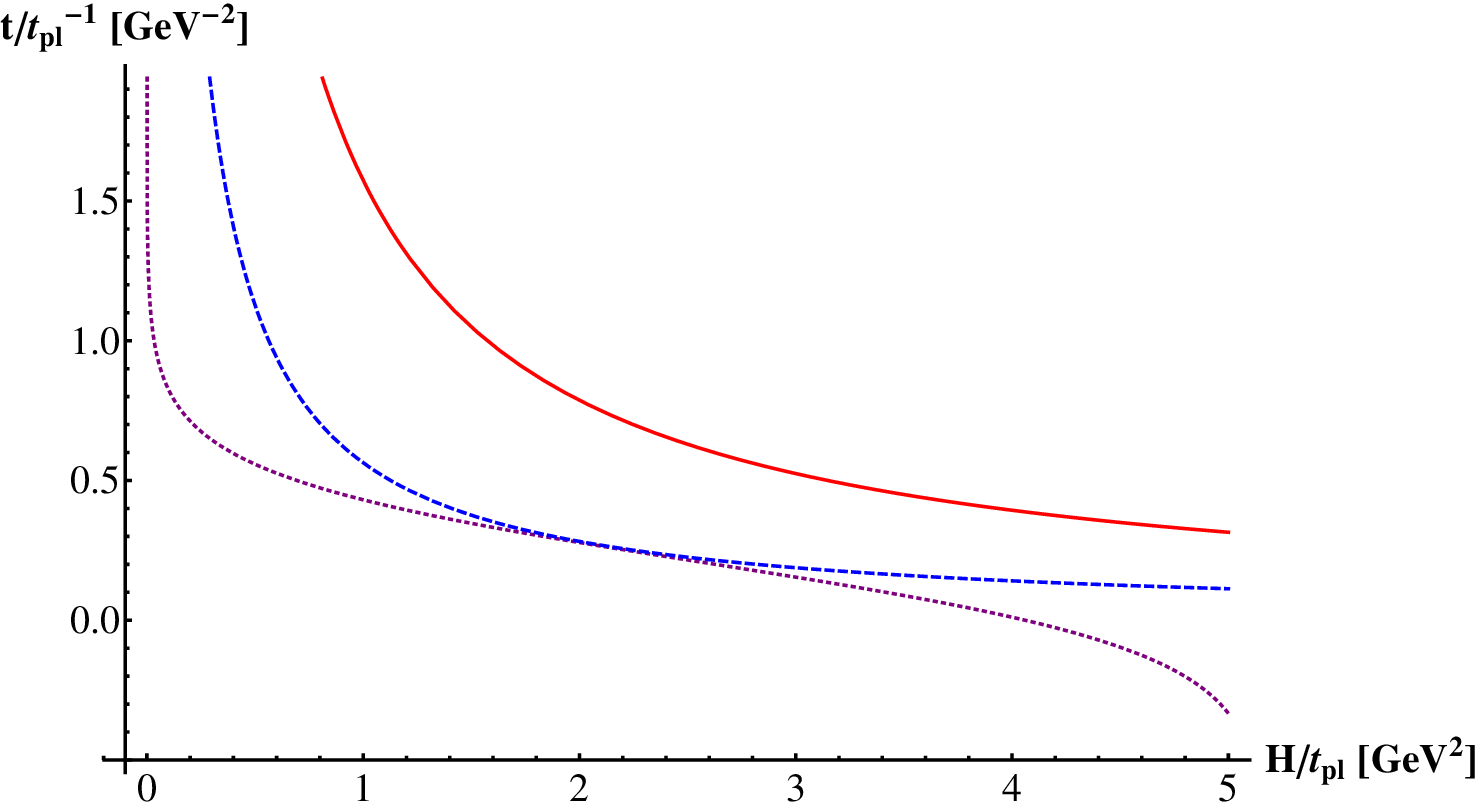}
\caption{The cosmological comoving  time $t$ vs Hubble parameter $H$, Eq. (\ref{eq:tHIS}), is graphically illustrated. The treatment of background matter is done by Israel-Stewart theory. The solid curve represents the results of present work, where the background matter is assumed to be characterized by viscous hadron and QGP. The dotted and dashed curves give the results when viscous QGP and an ideal (collisionless and nonviscous) gas, respectively, are assumed to fill the background geometry. Planck scale is given in physical units.
}
\label{afig:is-tH1}
\end{figure}

With the use of Eqs.~(\ref{aeq:p1}), (\ref{aeq:t}), (\ref{aeq:zeta2}) and (\ref{8}), respectively, we obtain the following equation describing the cosmological evolution of the Hubble parameter $H$: 
\bea
\ddot H +
\left\{\left(\frac{7}{2}-\frac{1}{2\pi}- \frac{3\alpha_1}{4\pi}\right) +\frac{3}{2\pi}
\left[\left(4\gamma_3 + \beta_2\,\gamma_4 \sqrt{\frac{6}{\pi}}\,H - 4\gamma_4\gamma_5\right)  \ln\left(\frac{\beta_2\,\sqrt{\frac{6}{\pi}}\,H - 4\gamma_1}{4 \gamma_2}\right)\right]^{-1}\,H\right\}\,H \dot H - & & \nonumber \\
   \frac{1}{4 \pi} \frac{\dot H^2}{H}+\frac{9}{4}(\alpha_1-1)\,H^3 +
\frac{9(1+\alpha_1)}{4\pi}
\left[\left(4\gamma_3 + \beta_2\,\gamma_4 \sqrt{\frac{6}{\pi}}\,H - 4\gamma_4\gamma_5\right)  \ln\left(\frac{\beta_2\,\sqrt{\frac{6}{\pi}}\,H - 4\gamma_1}{4 \gamma_2}\right)\right]^{-1}
\, H^4 &=&0. \hspace*{10mm} \label{eq:is1}
\eea
%
In obtaining this equation we have introduced a number of very tiny approximations. To the exponents $\alpha_2$, $\beta_3$ and $\gamma_6$ we have assigned the values, $1$, $1/2$ and $1$, respectively. In order to derive an analytical solution for this Abel differential equation, we follow the procedure given in \cite{twfk4,twfk8}. After some Algebra, it ends up with these two functions, 
\bea
g(H) &=& - \frac{4 \, \pi\, H^{-1-1/4\pi}}{14\, \pi -2-3\alpha_1 + 4\pi/\left\{\left[4\gamma_3+\beta_2\gamma_4\, \sqrt{6/\pi}\, H -4\gamma_4\gamma_5\right]\ln\left[\left(\beta_2\sqrt{6/\pi}\,H -4\gamma1\right)/4\gamma_2\right]\right\} }, \label{eq:isgH}\\
z(H) &=& \frac{H^{1-1/4\pi}}{(4\pi-1)(8\pi-1)\beta_2^2\gamma_4^2}\times\nonumber \\
&& \left\{4\pi(1-8\pi)\gamma_3 + 4\pi(8\pi-1)\left[\gamma_4\gamma_5+(\gamma_3-\gamma_4\gamma_5)\right] + \beta_2\gamma_4(4\pi-1)\left[\sqrt{6\pi}+(14\, \pi-2-3\alpha_1)\beta_2\gamma_4\right]H\right\}, \hspace*{10mm} \label{eq:iszH}
\eea
which are plotted in Fig.~(\ref{afig:is1}). They play an essential role in deriving an analytical solution for Eq. (\ref{afig:is1}). Approximating the parametric dependence of $g(H)$ on $z(H)$, Fig. \ref{afig:is-gz1}, we get the linear dependence, 
\bea
g(z) &\sim& 0.192\, z.
\eea
Then, from the definition of $\Omega$, we simply derive
\bea \label{OmegZ}
\Omega &=& - H^{-1/4\pi}\; \dot H.
\eea
In order to reduce this expression to the canonical equation of  Abel type, we use the relation $\Omega = z/{\cal P}$. Then from Eqs.~(\ref{OmegZ}) and (\ref{eq:iszH}), we obtain a first-order  differential equation for $H$,
\bea
{\cal P}\, \dot H &=& \frac{(4\pi-1)(8\pi-1)\beta_2^2\gamma_4^2}{H \left\{4\pi(1-8\pi)\gamma_3 + 4\pi(8\pi-1)\left[\gamma_4\gamma_5+(\gamma_3-\gamma_4\gamma_5)\right] + \beta_2\gamma_4(4\pi-1)\left[\sqrt{6\pi}+(14\, \pi-2-3\alpha_1)\beta_2\gamma_4\right]H\right\}}, \hspace*{10mm}
\eea
where ${\cal P}$ is a free parameter. The solution simply reads
\bea \label{eq:tHIS}
t&=& \frac{(8\pi -1) \beta_2 \gamma_4\; {\cal P}}{\left[\sqrt{6\pi} + \left(14\, \pi - 3 \alpha_1 -2\right) \beta_2 \gamma_4\right]\,H}.
\eea
The dependence of the cosmological comoving  time $t$ on the Hubble parameter $H$ is graphically illustrated in Fig. \ref{afig:is-tH1}. It is apparent that $t(H)$ is monotonic. The same dependence has been obtained, when assuming that the background matter is characterized as an ideal gas, $t=2/(3\gamma H)$. All this is summarized in Fig. \ref{afig:is-tH1}. Solid, dashed and dotted curves represent the results for viscous hadron-QGP, viscous QGP and ideal (nonviscous and collisionless) matter, respectively.

\section{Cosmological implications}\label{cosm1}

Assuming that the background geometry is filled with  Eckart relativistic {\it viscous} fluid, the comoving  time $t$ is given as a function of the Hubble parameter $H$ in  Eq.~(\ref{eq:eckrt1}) and drawn in Fig.~\ref{afig:eckart1} (solid curve). The dotted curve gives the results when {\it viscous} QGP EoS is implemented \cite{twfk1,twfk2,twfk3,twfk4,twfk5,twfk8,twfk9}. The dashed curve draws the results when the background geometry is assumed to be filled with an ideal gas. In Fig.~\ref{afig:is-tH1}, another $t$-$H$ dependence is obtained when assuming that the cosmological background is filled with  Israel-Stewart relativistic {\it viscous} fluid. The solid curve represents the results of present work, in which the background matter is assumed to be characterized by {\it viscous} hadrons and QGP i.e., including phase transition(s). The same treatment is applied for {\it viscous} QGP and ideal gas. The results are drawn by dotted and dashed curves, respectively.

Before discussing the potential cosmological implications, it is in order now to elaborate essential aspects. We start with the phase transition in the early universe. The first-order phase transition has been discussed in section \ref{SecII}. Section \ref{qgpEoS} was devoted to discuss the phase transition(s) as measured in lattice QCD simulations. Accordingly, we conclude that the order of the confinement-deconfinement phase transition depends among others on the effective degrees of freedom and the matter content (quark flavors, etc.) The results given in Figs. \ref{afig:eckart1} and \ref{afig:is-tH1} illustrate the effects of degrees of freedom (ideal gas, QGP and hadron-QGP matter) and in indirect way the phase transition (QGP matter above $T_c$ and hadron-QGP matter over a wide range of temperatures). The evolution of the Hubble parameter obviously depends on all these factors. This might have a direct cosmological implication that our picture about the expansion of the Universe has to be revised, accordingly.


Other cosmological implications might arise as a consequence of the phase transition itself. The first-order phase transition is to be characterized by a {\it sudden} change in the symmetry. It exhibits a {\it discontinuity} in the {\it first} derivative of the free energy with respect to some thermodynamic variable. In the cosmological context, such a transition is accompanied by bubble nucleation \cite{truner92}. In light of this, the Universe is conjectured to go from a metastable state to a new phase, a {\it true vacuum} state through the nucleation of bubbles of the new state \cite{truner92}. Implementing this model to the hadron-QGP transition makes it possible to suggest a scenario, in which the Universe starts from QGP state and ends up in the hadronic state through the nucleation of hadrons. Depending on the kinematics of the bubble nucleation, the Universe might or might not {''recover''} from this type of phase transition and  its relics are left behind i.e., relic QGP objects. The latter would survive for a very long time. The abundance and the size of the quark nuggets have been discussed in \cite{ind00}. Objects with a quark content ranging from $10^{-2}$ to $10$ $M_{\odot}$ could have been formed during the cosmological phase transition. 

Furthermore, a significant amount of entropy production is to be released during such a process, so that at vanishing chemical potential \hbox{$s=(c_s^2+1)\;\rho/T$}. The density fluctuations are assumed to be amplified by vanishing speed of sound during the quark-hadron phase transition, Fig. \ref{afig:1c}. The lattice QCD and hadron resonance gas calculations show that  the speed of sound reaches a minimum value, $c_s^2\simeq 0.1$, at $T_c$. On the other hand, the density fluctuations could produce QGP lumps decoupled from the expansion, which rapidly transform into quark nuggets. Typical distance between bubble centers is conjectured to be of the order of a few meters. It is worthwhile to mention here that the resulting baryon inhomogeneities may affect the primordial nucleosynthesis. Such a cosmological consequence can be observed. The origin of inhomogeneities in the matter distribution, which are assumed to be responsible for the later formation of galaxies, cannot be explained by density fluctuations, alone. After fixing the baryon number, the appearance of these fluctuations is almost purely adiabatic. Any departure from adiabaticity falling off is inversely proportional to the mass of the perturbation \cite{lindly_sci}. This will be elaborated in next paragraph.

At the phase transition, the scale of the cosmological QCD transition is assumed to be given by the Hubble radius $R_{H}$. Quantitatively, $R_{H}\simeq m_{Pl}/T_{c}^{2}\simeq 10\,$km. The mass inside the Hubble volume is $\simeq 1\,M_{\odot}$. At the QCD phase transition, the expansion time scale is $10^{-5}$ s, which is much large in comparison with the time-scale of QCD, $1$ fm/c $\simeq 10^{-23}\,$s. Even the rate of weak interactions seems to exceed the Hubble rate by a factor of $10^{7}$. Therefore, we conclude that photons (radiation), leptons, quarks (fermions) and gluons (bosons) are lightly coupled and may be described by an adiabatically expanding fluid \cite{twfk7,twfk9}, as the transition takes place  in an extremely short time. 

The current heavy-ion experiments program, LHC, seems to be very close to probe early eras of the universe. It seems to produce similar antiparticle and particle, when not entirely identical \cite{twfk10}. This can be taken as another supportive indicator for utilizing EoS deduced from heavy-ion collisions and/or lattice QCD calculations.  It seems that the observed matter-antimatter asymmetry can be explained without recourse to the hypothesis of specific initial conditions \cite{lindly_sci}.

\subsection{Different types of background matter}
\label{sec:mt}

The different phase transition(s) likely change the symmetries and thereupon different phases or types of matter are to be expected. 
The dynamics of the Universe during the fist-order phase transition from QGP to hadrons has been discussed in Section~\ref{SecIII}. Such a transition is assumed to go over three phases defined by various symmetries. {\it Prior} to the phase transition i.e., partonic (QGP) symmetry, the evolution of some cosmological parameters ($H$, $a$ and $\rho$) have been studied by using the Bianchi identity. To have an analytical insight into the evolution, $T$-corrections are neglected in the self-interaction potential. The second phase deals with the dynamics of the Universe {\it during} the phase transition i.e., mixed phase symmetry. Here, $T$ and $p$ are assumed to remain unchanged. The entropy $s$ and enthalpy $W$ remain conserved, as well. The third phase is the one in which the dynamics of the Universe is studied {\it post} quark-hadron phase transition era (hadronic symmetry). First we start with the time evolution of $T$. Then, we estimate the comoving  time $t$. Again, the Bianchi identity helps us in expressing the scale factor $a$ and the Hubble parameter $H$. The time evolution of the hadron fraction $h$ describes the conversion process of QGP into hadrons. Therefore, it can be taken as a parameter describing the phase transition itself.

Again, in this type of transition (a first-order phase transition through hadron nucleation), the numerical estimation of the cosmological parameters gives a clear indication that their time evolution varies from phase to another. In the QGP phase, the scale factor $a$ normalized to $a_0$ is much smaller than that in the hadronic phase, (compare Fig.~\ref{fig1} with the top panel in Fig.~\ref{fig3bb}). When studying the Hubble parameter $H$, the $T$-dependence is just the opposite of $a(T)$, (compare Fig.~\ref{fig2b} with the bottom panel in Fig.~\ref{fig3bb}). 

This behavior can be compared with the case of another types of phase transitions, {\it crossover }. In Figs.~\ref{afig:eckart1} and \ref{afig:is-tH1}, we notice that the time evolution of $H$ also depends on the type of matter filling the background geometry. If it is filled with QGP, the values of $H$ are relatively large. It is relatively small, if the background geometry is filled with quarks and hadrons, especially when  crossover phase transition is allowed to take place. Consequently, it is likely to predict that $H$ in the hadron era is smaller than its value in the previous eras: mixed phase of partons hadrons and QGP.

\begin{figure}[htb!]
\includegraphics[width=10.cm]{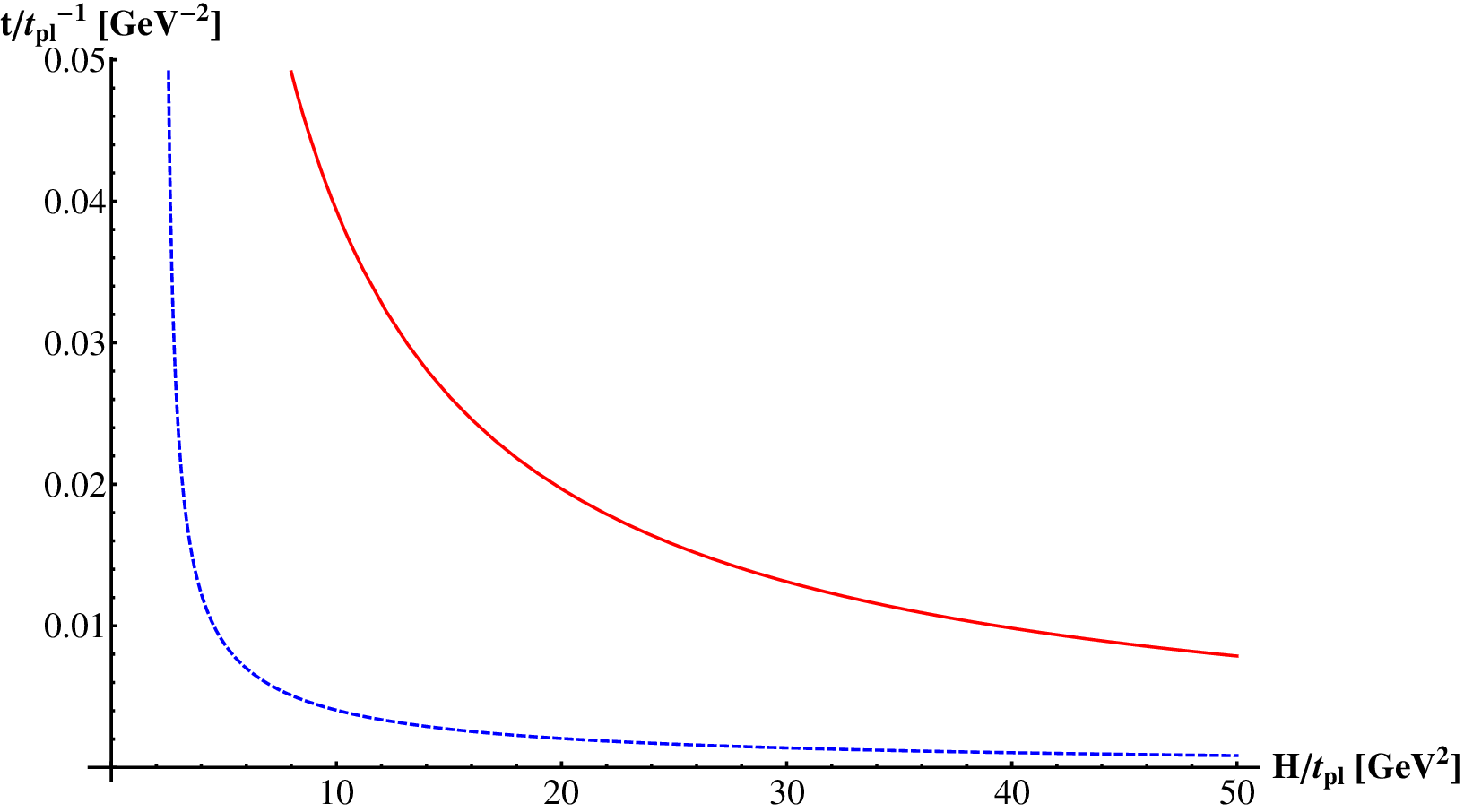}
\caption{The $t-H$ relation in Eckart relativistic QGP fluid. Solid and dotted curves represent viscous and nonviscous QGP fluid, respectively. It is obvious that the viscosity slows down the evolution of the Hubble parameter.
}
\label{afig:is-eckart-visocus}
\end{figure}

It seems to be in order now to highlight the differences between viscous and nonviscous background matter. By eliminating the dynamics controlling the phase transition, for instance, we assume that the background geometry is only filled with QGP. For simplicity, we utilize the Eckart theory. A comparison is illustrated in Fig. \ref{afig:is-eckart-visocus}. We notice that the viscosity seems to drastically slow down the evolution of the Hubble parameter. Should this result be confirmed, it would mean that the whole picture about the evolution of early Universe has to be revised. As a prompt consequence, one would expect a considerable delay in all phases post to QGP era. In order to make an estimation for this effect, other initial conditions have to be taken into consideration, for example, dynamics of phase transition(s), interaction(s), out-of-equilibrium processes, etc.

\section{Discussions and final remarks}\label{SecV}

In natural units, $\hbar=c=k_B=1$, all expressions are given the Planck mass $m_{pl}$. We consider the cosmic evolution of the early Universe in the regime of confinement QCD phase transition taking finite bulk viscous effects into account. Thereby, it is assumed that the bulk thermodynamic quantities are dominated by the strongly interacting matter component. Two cases, a first-order  phase transition scenario and an analytic crossover transition, are considered. In this respect, the present work continues a previous series \cite{twfk1,twfk2,twfk3,twfk4,twfk5,twfk6,twfk7,twfk8,twfk9} in several aspects. Refined equation(s) of state based on newer lattice QCD results are considered. Different bulk viscosity expressions based on quasiparticle model are used. Finite cosmological constant has been utilized in Ref. \cite{twfk3}. Moreover, the influence that a first-order  phase transition (neglecting viscous effects) is elaborated in the present work.

Many details of QCD phase transition(s) are not yet conclusively
understood. Even the order of transition is still a matter of
debate. An advance in understanding the numerical values of
the QCD coupling constants would be very helpful in obtaining
accurate cosmological conclusions \cite{twfk5}. Such an advance may also
provide a powerful method for testing on a cosmological scale the
theoretical predictions of the brane world models and the possible
existence of the extra-dimensions. Furthermore, the critical temperature $T_c$ has been a subject of different lattice QCD calculations~\cite{Fodor:2001au,deForcrand:2002ci,Allton:2002zi,D'Elia:2002gd,Karsch:2000kv,Karsch:2001cy,Gavai:2003mf}. In addition to this, it is still an open question whether both deconfinement and chiral phase transitions take place at the same $T_c$.

The cosmological behavior in first-order  phase transition can be characterized as follows. At the critical temperature, the energy $\rho$ and entropy $s$ densities decrease, suddenly. At fixed $T$ and constant $p$, both quantities have the same rate. Depending on the symmetries, the transition is assumed to go through three phases. {\it Prior} to the phase transition i.e., partonic (QGP) symmetry, the evolution of some cosmological parameters ($H$, $a$ and $\rho$) have been studied by using Bianchi identity. To have an analytical insight into the evolution, $T$ corrections are neglected in the self-interaction potential. The second phase is the one  during the phase transition i.e., mixed phase symmetry. Here, $T$ and $p$ are assumed to remain unchanged. Also the entropy $s$ and enthalpy $W$ remain conserved. The third phase is the one in which the dynamics of the Universe is studied {\it post} quark-hadron phase transition era i.e., hadronic symmetry. The Bianchi identity helps in expressing scale factor $a$ and Hubble parameter $H$. The behavior of $a$ and $H$ with the cosmological comoving times follows the standard cosmological model. Both quantities are expressed in terms of the fraction of matter. The latter gives an estimation for hadrons that are formed inside QGP. The time evolution of the hadron fraction describes the conversion process of QGP into hadrons. Therefore, it can be taken as a parameter describing the phase transition itself. A quantitative comparison between the evolution of scale factor $a$ in the three phases show that $a$ increases while moving from quarks to hadrons over the mixed phase. The values of the bag pressure are reflected in these calculations. In all phases we find that increasing the bag pressure raises the value of the scale factor.

Taking into account the recent lattice QCD results, we find that the order of the phase transition can be either continuous or discontinuous. It seems to depend on the quark flavors and their masses. The extreme conditions in the early universe, i.e., high temperatures, high densities and out-of-thermal and out-of-chemical equilibrium, likely affect the properties of the partonic matter and  control the dynamics of the phase transition. The equation of state deduced from lattice QCD calculations (and quasi particle model) plays a very essential role in present work. It sets the validity of the entire treatment. The high temperatures (energies), at which the strong coupling $\alpha_s$ nearly vanishes, defines the upper end of limitation. The lower one is characterized by the hadronic era. When applying Eckart theory, we find that the evolution of the Hubble parameter follows the same line defined by the standard cosmological model. The comparison with various types of matter shows that the comoving  time behaves very smooth with $H$, although viscous hadron-QGP results in larger $t$ than in viscous QGP. In both of them, $t$ seems to be larger than in the collisionfree and nonviscous {\it ideal} matter. Israel-Stewart theory is assumed to solve the constrains of Eckart theory. Therefore, reliable results are to be expected. In order to make a qualitative estimation for the effect of viscosity, we compare the time evolution of the Hubble parameter in a viscous and nonviscous background matter. Apparently, we find that the viscosity drastically slows down the evolution. Should this result be confirmed, the whole picture about the evolution of early Universe has to be revised, accordingly.  
In order to make an estimation for this effect, the dynamics of phase transition(s), interaction(s) and out-of-equilibrium and dissipative processes should be taken into account.
The effect of the cosmological constant on the anisotropy and homogeneity and the cosmological density perturbations in the early Universe would play an essential role in characterizing the evolution of the cosmological parameters  as well.

\section*{Acknowledgements}

The work of TH is supported by an RGC grant of the government of the Hong Kong SAR. The work of AT is partly supported by the German-Egyptian Scientific Projects (GESP ID: 1378).



\end{document}